\documentclass[11pt,a4paper]{article}

\usepackage[a4paper,
            bindingoffset=0in,
            left=0.9in,
            right=0.9in,
            top=0.9in,
            bottom=0.9in,
            footskip=.25in]{geometry}
\usepackage[utf8]{inputenc}
\usepackage{graphicx}
\usepackage{subfig}
\usepackage{placeins}
\usepackage{parskip}
\usepackage{amssymb}
\usepackage{amsmath}
\usepackage{bm}
\usepackage{amsthm}
\usepackage{algorithm}
\usepackage{algpseudocode}
\usepackage{comment}
\usepackage{hyperref}
\usepackage[sorting=nyt, style=authoryear, maxcitenames=2, uniquelist=false]{biblatex}
\DeclareNameAlias{sortname}{family-given}
\theoremstyle{plain}

\theoremstyle{definition}

\theoremstyle{remark}

\title{Bayesian Inference of Reproduction Number from Epidemiological and Genetic Data Using Particle MCMC}
\author{Alicia Gill\thanks{Nuffield Department of Medicine, University of Oxford, UK} \thanks{Corresponding author. Email: \href{mailto:alicia.gill@ndm.ox.ac.uk}{alicia.gill@ndm.ox.ac.uk}}, 
Jere Koskela\thanks{School of Mathematics, Statistics and Physics, Newcastle University, UK} \thanks{Department of Statistics, University of Warwick, UK}, 
Xavier Didelot\footnotemark[4] \thanks{School of Life Sciences, University of Warwick, UK} \thanks{The Zeeman Institute for Systems Biology and Infectious Disease Epidemiology Research, University of Warwick, UK}, 
Richard G. Everitt \footnotemark[4] \footnotemark[6]}
\date{}

\begin{document}

\maketitle

\FloatBarrier

\section*{Abstract}

\FloatBarrier

Inference of the reproduction number through time is of vital importance during an epidemic outbreak. Typically, epidemiologists tackle this using observed prevalence or incidence data. 
However, prevalence and incidence data alone is often noisy or partial. 
Models can also have identifiability issues with determining whether a large amount of a small epidemic or a small amount of a large epidemic has been observed. 
Sequencing data however is becoming more abundant, so approaches which can incorporate genetic data are an active area of research. 
We propose using particle MCMC methods to infer the time-varying reproduction number from a combination of prevalence data reported at a set of discrete times and a dated phylogeny reconstructed from sequences. 
We validate our approach on simulated epidemics with a variety of scenarios. 
We then apply the method to real data sets of HIV-1 in North Carolina, USA and tuberculosis in Buenos Aires, Argentina.
The models and algorithms are implemented in an open source R package called EpiSky which is available at \url{https://github.com/alicia-gill/EpiSky}.

\FloatBarrier

\section{Introduction}

\FloatBarrier

The basic reproduction number, denoted $R_{0}$, is a measure of the inherent transmissibility of a disease. 
It is defined as the average number of new infections caused by a single infected individual when entering a completely susceptible population (\cite{modernepi}). 
The effective or time-varying reproduction number, denoted $R_{t}$, represents the average number of new infections caused by a single infected individual at time $t$. 
The time-varying reproduction number is usually smaller than the basic reproduction number due to the depletion of susceptibles during the epidemic or the impact of control measures (\cite{wallinga2004}).
When $R_{t} < 1$, the epidemic is shrinking and when $R_{t}>1$, the epidemic is growing. 
The aim of infectious disease control programs is to reduce $R_{t}$ below 1. 
As such, estimating the reproduction number $R_{t}$ is of vital importance during an epidemic outbreak to decide which control measures to implement and to determine their effects once implemented.

Epidemiologists typically infer reproduction number $R_{t}$ using prevalence and/or incidence data (\cite{wallinga2004}, \cite{cori2013}, \cite{boon2022}). 
However, data relating to cases is often noisy, partially observed, and biased. 
It can also be difficult to identify whether there have been many cases of a small epidemic observed or few cases of a large epidemic. 
In fact, modelling approaches to find the reporting probability of an epidemic is an active area of research (\cite{kokouvi2014}, \cite{kokouvi2017}, \cite{stoner2019}, \cite{liang2022}). 
Given these challenges with epidemiological data, genetic data, typically in the form of a dated phylogeny (see Section \ref{subsec:phylo}), is increasingly being used to understand infectious disease epidemiology (\cite{rasmussen2011}, \cite{li2017}, \cite{vaughan2019}).

To combine partial epidemiological information with genetic data, we use a Bayesian approach to find the posterior distribution of $R_{t}$ given the partially observed epidemic and the genetic data. 
We then use pseudo-marginal MCMC (\cite{andrieu2009}) with likelihoods estimated using sequential Monte Carlo (SMC) (\cite{chopin}) to infer $R_{t}$ trajectories. 
Specifically, we will use the particle marginal Metropolis--Hastings (PMMH) algorithm (\cite{andrieu2010}) to propose parameters and latent prevalence trajectories.

There are a few methods using PMMH to combine prevalence data with (phylo)genetic data, most notably \textcite{rasmussen2011} and \textcite{judge2024}. 
\textcite{judge2024} use birth-death phylogenetic trees from \textcite{stadler2009} and \textcite{stadler2010} whereas we use coalescent tree models.
\textcite{rasmussen2011} also assume coalescent phylogenetic trees.
A key difference with our method is that we suppose that the prevalence and genetic data are related to each other, rather than independent of each other. 
There is only pseudo-code associated with \textcite{rasmussen2011}.
We have developed an R package to implement our model called \textit{EpiSky}, available at \url{https://github.com/alicia-gill/EpiSky}, where the PMMH algorithm can be implemented for model fitting.
We have also built in considerable adaptation into our procedures to avoid computationally expensive hand-tuning of MCMC.

\subsection{Phylogenetics} \label{subsec:phylo}

A phylogenetic tree or phylogeny is a diagram that represents evolutionary history. 
For an infectious disease, leaves denote sampled pathogens and internal nodes denote most recent common ancestors between those pathogens. 
In order to use phylogenies for epidemiology, we need them to be scaled according to calendar time, i.e. a dated phylogeny.
In a dated phylogeny, leaves are aligned with the known date of isolation of the genomes and nodes are aligned with the estimated dates of common ancestors. 
Many methods exist to reconstruct dated phylogenies from sampled sequence data, including BEAST (\cite{beast}), BEAST2 (\cite{beast2}), LSD (\cite{lsd}), BactDating (\cite{bactdating}), treedater (\cite{treedater}) and TreeTime (\cite{treetime}).

A common analytical approach in the literature when using phylogenetic trees for inference is to first reconstruct a dated phylogenetic tree from sequencing data and to then draw some epidemiological interpretation from it (\cite{didelot2022}). 
This is the approach that we take. 
Whilst this does not fully incorporate the uncertainty in the phylogenetic tree into the modelling, it does allow for considerable scalability, thus allowing us to take advantage of the abundance of genetic data.

We use coalescent theory to model the phylogeny (\cite{kingman1982a}) extended to allow for heterochronous sampling and a natural time scale (\cite{drummond2002}). 
The coalescent is robust to temporal bias in the genetic sampling as it models the ancestry given the sampling dates (\cite{volz2014}). 
This is unlike methods modelling phylogenies as birth-death processes, as these model both the branching and the sampling (\cite{stadler2009}, \cite{stadler2010}). 
Unlike the epidemiological data for which we will assume the reporting probability is constant through time, there is no such assumption for the genetic data.

Several methods exist to estimate the effective population size over time given a dated tree (\cite{ho2011}; \cite{karcher2017}; \cite{volz2018}; \cite{didelot2023}). 
However, this is not the same as estimating $R_{t}$. There is a complex relationship, not simply a proportional one, between the effective population size and the prevalence (\cite{volz2009}; \cite{frost2010}; \cite{volz2012}). 
We will discuss how to relate the effective population size to $R_{t}$ in Section \ref{subsec:genlik}.

\FloatBarrier

\section{Bayesian modelling}

\FloatBarrier

\subsection{Latent epidemic model}

Let $\{X_{t}\}_{t \geq 0}$ denote the number of infected individuals at time $t$. 
We can model this epidemiological process as a non-homogeneous birth-death process (\cite{kendall1948}). 
Modelling the epidemic in this way is akin to a susceptible-infectious-susceptible compartmental model with infinite population, so has been chosen for its simplicity and convenience.

The transition probabilities are given by
\begin{equation}
\label{eq:bdp}
\begin{split}
&\mathbb{P}(X_{t+h} - X_{t} = 1 \mid X_{t} = i) = \beta_{t}ih + o(h), \\
&\mathbb{P}(X_{t+h} - X_{t} = -1 \mid X_{t} = i) = \gamma_{t}ih + o(h), \\
&\mathbb{P}(|X_{t+h} - X_{t}| > 1 \mid X_{t} = i) = o(h), \\
&\mathbb{P}(X_{t+h} - X_{t} = 0 \mid X_{t} = i) = 1 - (\beta_{t} + \gamma_{t})ih + o(h),
\end{split}
\end{equation}
where $h \to 0$ is a small interval, $\beta_{t}$ is the birth rate and $\gamma_{t}$ is the death rate at time $t$. 
We will suppose that the death rate $\gamma_{t} \equiv \gamma$ is constant and known. 
Supposing that the death rate is known is a simplifying assumption to circumvent potential non-identifiability between the birth rates and the death rates.
Having a constant, known death rate is often not unreasonable in practice, as we are likely to have data on the average length of infection $1/\gamma$.


In a birth-death model of disease outbreak,
\begin{equation}
R_{t} = \frac{\beta_{t}}{\gamma_{t}} = \frac{\beta_{t}}{\gamma}.
\end{equation}
Therefore, inferring $\beta_{t}$ will allow us to infer $R_{t}$. 

Although the underlying epidemic process is a continuous-time Markov process, we will only be observing it in discrete time, as we will have access to the prevalence for each unit of time. 
Note that these time units could be any arbitrary time step, but we will refer to one unit of time as a day throughout. 
In order to match the data, we will approximate the continuous-time process in discrete time.
Let $\{X_{n}\}_{n \in \mathbb{N}}$ denote this discretised process. 
In discrete time, for an epidemic that has been ongoing for $N$ days, we are interested in modelling the discrete transition probability, 
\begin{equation}
p(X_{1:N} \mid X_{0}, \beta_{1:N}, \gamma) = \prod_{n=1}^{N} p(X_{n} \mid X_{n-1}, \beta_{n}, \gamma).
\end{equation}
Using rates given by supposing that the transition probabilities are constant over one unit of time $h$ in equation (\ref{eq:bdp}), we model birth events on day $n$ as Poisson distributed with rate $\beta_{n}x_{n-1}$ and death events as Poisson distributed with rate $\gamma x_{n-1}$ when there are $x_{n-1}$ infectious individuals on day $n-1$. 
Then the prevalence on day $n$ is the prevalence on day $n-1$ plus the number of births on day $n$ minus the number of deaths on day $n$.
This is a natural and statistically convenient way to approximate the continuous-time birth-death process in discrete time. 
This approximation is best when there are only small changes in the prevalence between days. 
The difference between two (conditionally independent) Poisson distributions is given by the Skellam distribution (\cite{skellam}), so we use this to calculate $p(X_{1:N} \mid X_{0}, \beta_{1:N}, \gamma)$.

Let $B_{n}$ be the random variable representing the birth events on day $n$ and let $D_{n}$ be the random variable representing the death events on day $n$. 
Then $B_{n} \mid (X_{n-1}=x_{n-1}, \beta_{n}) \sim \text{Poisson}(\beta_{n} x_{n-1})$ and $D_{n} \mid (X_{n}=x_{n}, \gamma) \sim \text{Poisson}(\gamma x_{n-1})$, so $B_{n}-D_{n} \mid (X_{n-1}=x_{n-1}, \beta_{n}, \gamma) \sim \text{Skellam}(\beta_{n} x_{n-1}, \gamma x_{n-1})$. 
Note that $B_{n}$ and $D_{n}$ are conditionally independent given $x_{n-1}$. 
The likelihood for the latent epidemic is:
\begin{equation}
\begin{split}
p(X_{1:N} = x_{1:N} \mid X_{0} = x_{0}, \beta_{1:N}, \gamma) 
&= \prod_{n=1}^{N} \mathbb{P}(X_{n}=x_{n} \mid X_{n-1}=x_{n-1}, \beta_{n}, \gamma) \\
&= \prod_{n=1}^{N} \mathbb{P}(B_{n}-D_{n}=x_{n}-x_{n-1} \mid X_{n-1}=x_{n-1}, \beta_{n}, \gamma) \\
&= \prod_{n=1}^{N} e^{-(\beta_{n}+\gamma)x_{n-1}} \left( \frac{\beta_{n}}{\gamma} \right)^{\frac{x_{n}-x_{n-1}}{2}} I_{|x_{n}-x_{n-1}|} (2x_{n-1}\sqrt{\beta_{n}\gamma})
\end{split}
\end{equation}
where $I$ is the modified Bessel function of the first kind.

This likelihood should ideally be conditioned to assign probability 1 to the event that the epidemic does not die out or become negative, i.e. $B_{n} - D_{n} \mid (X_{n-1}=x_{n-1}, \beta_{n}, \gamma) \sim \text{Skellam}(\beta_{n} x_{n-1}, \gamma x_{n-1}) \mid (B_{n}-D_{n} > -x_{n-1})$. 
However, this conditioning is computationally expensive and the probability of sampling below $-x_{n-1}$ is small, especially as the epidemic grows. 
As such, we omit the condition in practice.

\subsection{Epidemic observation model}

Let $Y_{n}$ denote the observed prevalence on day $n$. 
We next consider how to model this observed epidemic $Y_{1:N}$ given the latent epidemic $X_{1:N}$ and the reporting probability $\rho$, $p(Y_{1:N} \mid X_{1:N}, \rho) = \prod_{n=1}^{N} p(Y_{n} \mid X_{n}, \rho)$. 
There are many possible ways to model observed prevalence given latent prevalence depending on what is believed to be observed, for example, complete prevalence data with noise or partial cases. 
We will suppose that we are observing some proportion $\rho$ of cases at random, i.e. $Y_{n} \mid X_{n}=x_{n}, \rho \sim \text{Binomial}(x_{n}, \rho)$ $\forall n = 1, \ldots, N$. 
$X_{n}$ and $\rho$ are independent a priori.

\FloatBarrier
\subsection{Phylogenetic model} \label{subsec:genlik}
\FloatBarrier

Lastly, we consider how to model the dated phylogeny, inferred from sequencing data, given the birth rates and latent prevalence. 
We model the phylogeny as the realisation of a random tree with a fixed set of leaves situated at the known sampling times. 
The tree is then generated by tracing lineages starting from the leaves backwards in time and merging each contemporaneous pair at rate $\lambda_{t}$. 
In a Kingman's coalescent, the (exponential) rate at which these coalescence events occur is given by (\cite{koelle2012})
\begin{equation}
\lambda_{t} = \binom{A_{t}}{2} \frac{1}{N_{e}(t)}
\end{equation}
where $A_{t}$ denotes the number of lineages and $N_{e}(t)$ denotes the effective population size at time $t$. 
The effective population size reflects the number of individuals that contribute offspring to the descendent generation and is almost always smaller than the census population size (\cite{ho2011}). 

\textcite{volz2009} derived a relationship between the coalescence rate in a dated phylogeny and the transmission rate in a birth-death or SIR (susceptible-infectious-recovered) epidemic. 
Assuming that superinfection is rare and that mutation is fast relative to epidemic growth, each lineage in a phylogenetic tree corresponds to a single infected individual with its own unique viral population.
Then backwards in time, an infection event in the epidemic is equivalent to a coalescent event in the phylogeny, and forwards in time, each transmission of the pathogen between hosts can generate a new branch in the phylogeny of consensus isolates from infected individuals.
New cases correspond to transmission events in the epidemic. 
However, not all transmission events in the epidemic will result in an observed coalescence event in the phylogeny.
For a transmission event to show in the phylogeny requires that both the infector and the infectee are sampled.
This sampling can be due to either the individuals being sampled or their descendant lineages being sampled.
Given that a coalescence event occurs among the $X_{t}$ infected individuals at time $t$, the probability of observing it among the $A_{t}$ observed lineages at time $t$ is
\begin{equation}
\label{eqn:trans_rate}
\frac{\binom{A_{t}}{2}}{\binom{X_{t}}{2}}.
\end{equation}
Let $f_{SI}(t)$ denote the rate at which susceptible individuals become infected at time $t$.
Then the coalescence rate $\lambda_{t}$ at time $t$ is the transmission rate $f_{SI}(t)$ multiplied by the probability of observing that event $\binom{A_{t}}{2}/\binom{X_{t}}{2}$:
\begin{equation}
\label{eqn:trans_rate_deriv}
\lambda_{t} 
= \binom{A_{t}}{2} \frac{f_{SI}(t)}{\binom{X_{t}}{2}}
\approx \binom{A_{t}}{2} \frac{2f_{SI}(t)}{X_{t}^{2}}.
\end{equation}

In a birth-death model of disease outbreak, the transmission rate $f_{SI}(t) = \beta_{t}X_{t}$, so
\begin{equation}
\lambda_{t} \approx \binom{A_{t}}{2} \frac{2\beta_{t}}{X_{t}}.
\label{eqn:coal}
\end{equation}

\begin{figure}[ht]
\centering
\subfloat[][]{%
\includegraphics[width=0.9\textwidth]{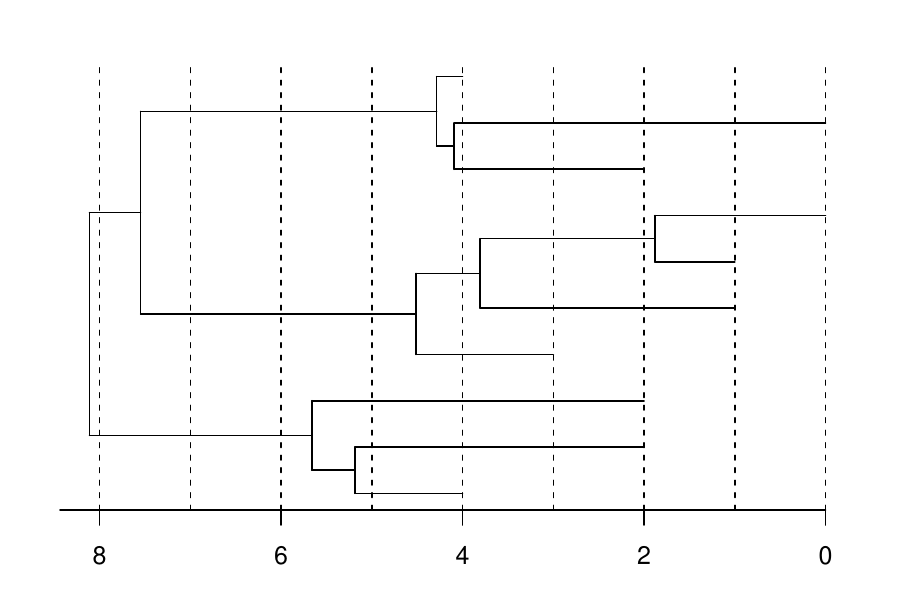}
}%
\vfill
\subfloat[][]{%
\begin{tabular}{r|ccccccccc}
Days from the present & 8 & 7 & 6 & 5 & 4 & 3 & 2 & 1 & 0 \\ 
$A_{n}$               & 2 & 3 & 3 & 5 & 8 & 7 & 6 & 4 & 2 \\ 
$C_{n}$               & 1 & 1 & 0 & 2 & 3 & 1 & 0 & 1 & 0 \\ 
\end{tabular}
}%
\caption{(a) A phylogenetic tree segmented by days from the present. (b) A table showing the corresponding values of $A_{n}$ and $C_{n}$}
\label{fig:disctree}
\end{figure}

Since $X_{t}$ is a process we infer in discrete time, it is therefore natural to also segment the dated phylogeny (see Figure \ref{fig:disctree}). 
We do this by slicing the phylogeny into days, starting with the present day, and going backwards in time. On each day $n$, we have some number of lineages denoted $A_{n}$ and some number of coalescence events $C_{n}$. 

From (\ref{eqn:coal}) we know that coalescence events on day $n$ occur at rate $\binom{A_{n}}{2} 2\beta_{n} / X_{n}$. 
This means that each pair of lineages $A_{n}$ coalesce at (exponential) rate $2 \beta_{n} / X_{n}$. 
We model the number of coalescences as a binomial distribution with $\binom{A_{n}}{2}$ trials and success probability $1 - \exp(-2\beta_{n}/X_{n})$. 
The likelihood for the dated phylogeny is therefore:

\begin{equation}
\begin{split}
p(G_{1:N} \mid \beta_{1:N}, X_{1:N}) 
&= \prod_{n=1}^{N} p(G_{n} \mid \beta_{n}, X_{n}) \\
&= \prod_{n=1}^{N} p(C_{n} = c_{n} \mid A_{n}=a_{n}, \beta_{n}, X_{n}=x_{n}) \\
&= \prod_{n=1}^{N} \binom{\binom{a_{n}}{2}}{c_{n}} \left(1 - \exp\left(-\frac{2\beta_{n}}{X_{n}}\right) \right)^{c_{n}} \exp\left(-\frac{2\beta_{n}}{X_{n}}\right) ^ {\binom{a_{n}}{2} - c_{n}}.
\end{split}
\end{equation}

\FloatBarrier
\subsection{State space model}
\FloatBarrier

\begin{figure}[ht]
\centering
\includegraphics[width=0.9\textwidth]{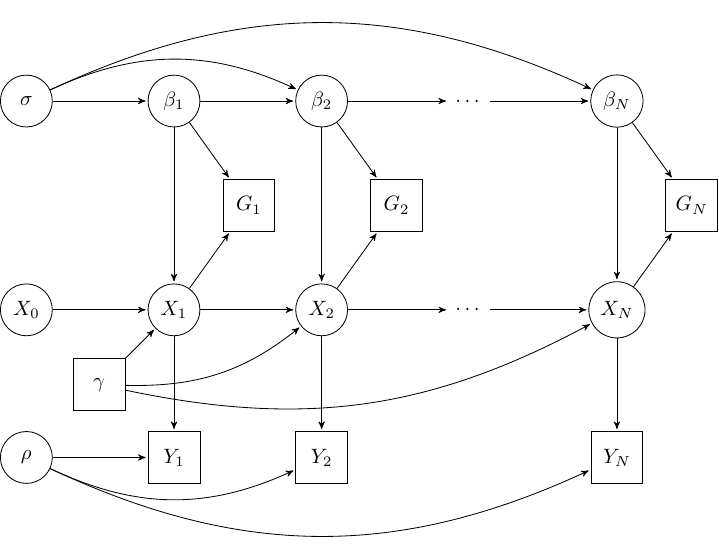}
\caption{Bayesian network representing the conditional dependencies between the birth rates $\beta_{n}$, true prevalence $X_{n}$, genetic data $G_{n}$, observed prevalence $Y_{n}$ on days $n=\{1,\ldots,N\}$, the death rate $\gamma$ and parameters $\sigma$ and $\rho$. Square nodes denote observed variables and circular nodes denote unobserved variables.}
\label{fig:hmm}
\end{figure}

Suppose the epidemic has been ongoing for $N$ days. 
The aim is to infer the unknown birth rate trajectory $\beta_{1:N}$ given the segmented dated phylogeny $G_{1:N}$ and the partially observed epidemic $Y_{1:N}$. 
Since the epidemic has been observed discretely and we are discretising all other processes, we can present this in a state space model framework as seen in Figure \ref{fig:hmm}.

Let $\beta_{n}$ denote the birth rate on day $n$, $G_{n}$ denote the slice of the phylogenetic tree from day $n-1$ to $n$, $X_{n}$ denote the latent prevalence on day $n$ and $Y_{n}$ denote the observed prevalence on day $n$. 
Let $\sigma$ denote the amount that the birth rate trajectory is allowed to vary between days (details in Section $\ref{subsec:priors}$) and let $\rho$ denote the reporting probability, that is, the proportion of cases that have been observed.

We adopt a Bayesian approach to find $p(\beta_{1:N}, \sigma, \rho, X_{0} \mid \gamma, G_{1:N}, Y_{1:N})$. 
\begin{equation}
\label{eqn:likelihood}
\begin{split}
&p(\beta_{1:N}, \sigma, \rho, X_{0} \mid \gamma, G_{1:N}, Y_{1:N}) \\
&\propto p(\beta_{1:N}, \sigma, \rho, X_{0}, \gamma, G_{1:N}, Y_{1:N}) \\
&= \int p(\beta_{1:N}, \sigma, \rho, X_{0}, \gamma, G_{1:N}, Y_{1:N}, X_{1:N}) \, dX_{1:N} \\
&= \underbrace{p(\sigma) p(\rho) p(X_{0})}_{\text{parameters}}
\underbrace{p(\beta_{1:N} \mid \sigma)}_{\text{birth rates}}
\int 
\underbrace{p(X_{1:N} \mid X_{0}, \beta_{1:N}, \gamma)}_{\text{latent epidemic}}
\underbrace{p(G_{1:N} \mid \beta_{1:N}, X_{1:N})}_{\text{phylogeny}}
\underbrace{p(Y_{1:N} \mid X_{1:N}, \rho)}_{\text{observed epidemic}}
\, dX_{1:N} \\
\end{split}
\end{equation}

The problem is that to evaluate this explicitly requires integrating over the space of all possible epidemics, which grows in the length of the time series. 
This is not analytically tractable. We will discuss methods to tackle this problem in Section \ref{sec:PMMH}. 
The rest of this section will present the models used to relate the data and the parameters.

\FloatBarrier

\section{Particle MCMC} \label{sec:PMMH}

\FloatBarrier

\subsection{Pseudo-marginal MCMC}

As mentioned in the previous section, the distribution that we wish to target, $p(\beta_{1:N}, \sigma, \rho, X_{0} \mid \gamma, G_{1:N}, Y_{1:N})$, is intractable. 
This prevents us from directly using standard likelihood-based inference tools, such as the Metropolis--Hastings algorithm (\cite{hastings1970}), as we cannot evaluate the likelihood for each value of parameters $\sigma$, $\rho$ and $X_{0}$.

One approach to get around this could be data-augmented MCMC, where we augment the state space with the latent prevalence $X_{1:N}$ and the birth rate trajectory $\beta_{1:N}$ (\cite{vandyk2001}). 
However, since we are considering time series data, this method would vastly increase the number of variables over which to explore. 
Moreover, these variables would be highly correlated as future days in the time series depend upon previous ones. 
As such, these chains would be difficult to tune and mixing would likely be poor, and thus would need to be run for a very long time.

Instead we turn to a different class of MCMC methods called pseudo-marginal methods (\cite{andrieu2009}). 
In pseudo-marginal MCMC, we use an unbiased estimator of the analytically intractable likelihood, $\hat{p}(\beta_{1:N}, \sigma, \rho, X_{0} \mid \gamma, G_{1:N}, Y_{1:N})$, in place of the true likelihood, $p(\beta_{1:N}, \sigma, \rho, X_{0} \mid \gamma, G_{1:N}, Y_{1:N})$, in the Metropolis--Hastings algorithm. 
We do this by simulating several possible $X_{1:N}$ and $\beta_{1:N}$ for fixed parameters $\sigma$, $\rho$ and $X_{0}$, and then using these samples to estimate the likelihood of the data given these samples. 
We will derive this estimated likelihood using SMC, so specifically we will use the PMMH algorithm (\cite{andrieu2010}).

We target an acceptance rate of 10\% using the adaptive scaling within adaptive Metropolis algorithm from \textcite{vihola2011}. 
This is close to optimal values demonstrated in \textcite{sherlock2015}.

\subsection{Sequential Monte Carlo}
\label{subsec:pf}

It is natural with time series data to sample sequentially in order to best exploit the structure of the data. 
For example, we know that the prevalence on day $n$ depends on the prevalence on day $n-1$, so we would like to sample the prevalence on day $n$ conditional upon the prevalence on day $n-1$. 
We can take advantage of this sequential sampling if on each day $n$ we resample in order to keep the most likely proposals. 
This idea is the basis for SMC algorithms (\cite{doucet2011}).
The SMC algorithm we use is shown in Algorithm \ref{alg:smc}.

\begin{algorithm}[ht]
\caption{SMC}
\label{alg:smc}
\begin{algorithmic}
\State \textbf{Input:} Values for parameters $\theta=(\sigma, \rho, X_{0})$, number of particles $K$, death rate $\gamma$, observed prevalence data $Y_{1:N}$, discretised phylogenetic tree $G_{1:N}$
\State \textbf{Output:} $L^{K}$ - an estimator for $\int p(\beta_{1:N}, X_{1:N}, G_{1:N}, Y_{1:N} \mid \theta, \gamma) \, dX_{1:N}$
\For{$n = 1, \ldots, N$}
    \State Draw $(\beta_{n}^{k}, X_{n}^{k}) \sim q_{\theta}(\cdot \mid \beta_{1:n-1}^{k}, X_{1:n-1}^{k})$ for $k = 1, \ldots, K$
    \State Weight the pairs $(\beta_{n}^{k}, X_{n}^{k})$ as 
        \begin{equation*}
            w_{n}^{k} 
            = \frac{p_{\theta}(\beta_{1:n}^{k}, X_{1:n}^{k}, G_{1:n}, Y_{1:n})}{p_{\theta}(\beta_{1:n-1}^{k}, X_{1:n-1}^{k}, G_{1:n-1}, Y_{1:n-1}) q_{\theta}(\beta_{n}^{k}, X_{n}^{k} \mid \beta_{1:n-1}^{k}, X_{1:n-1}^{k})}
        \end{equation*}
    \State Normalise $W_{n}^{k} = w_{n}^{k} / \sum_{j=1}^{K} w_{n}^{j}$
    \State Resample ancestors $A_{n}^{1:K}$ according to the normalised weights and keep pairs $(\beta_{n}^{A_{n}^{1:K}}, X_{n}^{A_{n}^{1:K}})$
    \State Set $l_{n}^{K} = \frac{1}{K} \sum_{k=1}^{K} w_{n}^{k}$
    \State Set $w_{n}^{k} = 1$
\EndFor \\
\State Set $L_{n}^{K} = \prod_{t=1}^{n} l_{t}^{K}$
\Return $L_{N}^{K}$
\end{algorithmic}
\end{algorithm}

The resampling step will be discussed in more detail in Section \ref{subsec:resampling}. 
If desired, we can also output samples of trajectories $\beta_{1:N}$ and/or $X_{1:N}$ by choosing a trajectory to output according to its weight at step $N$.

\subsection{Particle marginal Metropolis--Hastings}

Let us denote by $\hat{p}_\mathrm{SMC}$ the estimator for the likelihood from the SMC algorithm. 
The PMMH algorithm is a specific kind of pseudo-marginal sampler where the unbiased estimator used is $\hat{p}_\mathrm{SMC}$.
This is shown in Algorithm \ref{alg:pmmh}.

\begin{algorithm}[ht]
\caption{PMMH}
\label{alg:pmmh}
\begin{algorithmic}
\State \textbf{Input:} Initial values $\theta^{(0)}$ for unknown parameters ($\sigma^{(0)}$, $\rho^{(0)}$, $X_{0}^{(0)}$), number of iterations $I$, number of particles $K$, death rate $\gamma$, observed prevalence data $Y_{1:N}$, phylogenetic tree $G$
\State \textbf{Output:} $\bm{\sigma}$ - a vector of length $I$, $\bm{\rho}$ - a vector of length $I$, $\bm{X_{0}}$ - a vector of length $I$, $\bm{\beta}$ - a matrix with dimension $I \times N$
\State Run the SMC algorithm targeting $f(\beta_{1:N}^{(0)}, X_{1:N}^{(0)} \mid \theta^{(0)}, \gamma, Y_{1:N}, G)$, simulate $K$ pairs of trajectories $(\beta_{1:N}^{(0)}, X_{1:N}^{(0)})$ and compute $\hat{p}_\mathrm{SMC}(\beta_{1:N}^{(0)}, X_{1:N}^{(0)} \mid \theta^{(0)}, \gamma, Y_{1:N}, G)$
\For{$i = 1, \ldots, I$}
    \State Draw $\theta^{*} \sim q(\cdot \mid \theta^{(i-1)})$
    \State Run the SMC algorithm targeting $f(\beta_{1:N}^{*}, X_{1:N}^{*} \mid \theta^{*}, \gamma, Y_{1:N}, G)$, simulate $K$ pairs of trajectories $(\beta_{1:N}^{*}, X_{1:N}^{*})$ and compute $\hat{p}_\mathrm{SMC}(\beta_{1:N}^{*}, X_{1:N}^{*} \mid \theta^{*}, \gamma, Y_{1:N}, G)$
    \State Compute the ratio $r = \frac{ p(\theta^{*}) \hat{p}_\mathrm{SMC}(\beta_{1:N}^{*}, X_{1:N}^{*} \mid \theta^{*}, \gamma, Y_{1:N}, G) q(\theta^{(i-1)} \mid \theta^{*}) }{ p(\theta^{(i-1)}) \hat{p}_\mathrm{SMC}(\beta_{1:N}^{(i-1)}, X_{1:N}^{(i-1)} \mid \theta^{(i-1)}, \gamma, Y_{1:N}, G) q(\theta^{*} \mid \theta^{(i-1)}) }$
    \State Draw $u \sim \text{Uniform}(0,1)$
    \If{$u \leq \min\{1, r \}$}
        \State $\theta^{(i)} = \theta^{*}$
        \State $(\beta_{1:N}^{(i)}, X_{1:N}^{(i)}) = (\beta_{1:N}^{*}, X_{1:N}^{*})$
        \State $\hat{p}_\mathrm{SMC}(\beta_{1:N}^{(i)}, X_{1:N}^{(i)} \mid \theta^{(i)}, \gamma, Y_{1:N}, G) = \hat{p}_\mathrm{SMC}(\beta_{1:N}^{*}, X_{1:N}^{*} \mid \theta^{*}, \gamma, Y_{1:N}, G)$
    \Else
        \State $\theta^{(i)} = \theta^{(i-1)}$
        \State $(\beta_{1:N}^{(i)}, X_{1:N}^{(i)}) = (\beta_{1:N}^{(i-1)}, X_{1:N}^{(i-1)})$
        \State $\hat{p}_\mathrm{SMC}(\beta_{1:N}^{(i)}, X_{1:N}^{(i)} \mid \theta^{(i)}, \gamma, Y_{1:N}, G) = \hat{p}_\mathrm{SMC}(\beta_{1:N}^{(i-1)}, X_{1:N}^{(i-1)} \mid \theta^{(i-1)}, \gamma, Y_{1:N}, G)$
    \EndIf
    \State $\bm{\sigma}[i,] = \sigma^{(i)}$
    \State $\bm{\rho}[i,] = \rho^{(i)}$
    \State $\bm{X_{0}}[i,] = X_{0}^{(i)}$
    \State Sample one trajectory $\beta_{1:N}^{k}$ from $\beta_{1:N}^{(i)}$
    \State $\bm{\beta}[i,] = \beta_{1:N}^{k}$
\EndFor \\
\Return $\bm{\sigma}$, $\bm{\rho}$, $\bm{X_{0}}$, $\bm{\beta}$
\end{algorithmic}
\end{algorithm}

\subsubsection{Proposal mechanisms}

The birth rates are sampled according to their prior. 
That is, the birth rate on the first day is supposed to be exponential with rate $(2\gamma)^{-1}$ and on subsequent days are normally distributed centred on the previous birth rate with variance $\sigma^{2}$.

Values for the latent prevalence are sampled from a mixture distribution of the Skellam prior and a proposal distribution informed by the data. 
The balance of the mixture is chosen depending on the reporting probability $\rho$. 
Given $\rho$, we propose $x_{n} \sim p \cdot \text{NegBin}(y_{n}, \rho) + (1-p) \cdot \text{Skellam}(\beta_{n} x_{n-1}, \gamma x_{n-1})$. 
In practice, we say $p = \min\{ \rho/0.1 , 0.95 \}$, which gives a mixture proposal that is very data driven, but also allows the samples to take reasonable values when in the tails of the reporting probability space (i.e. when the data-driven proposals would be misleading).

\subsection{Path degeneracy} \label{subsec:pathdegen}

As the length of the epidemic increases (i.e. as $N$ increases), the resampling step in the SMC algorithm necessarily results in a phenomenon called path degeneracy, where all trajectories with high weights on day $N$ start from one sample on days $1, \ldots, t$ for some $t<N$ (\cite{doucet2011}). 
This path degeneracy can lead to poor inference of the birth rate for earlier days of the epidemic. 
There are a few techniques to combat this path degeneracy which we will discuss in the rest of this section.

\subsubsection{Resampling schemes} \label{subsec:resampling}

There are several options for the resampling step in the SMC algorithm, as the only constraint is that the resampling mechanism gives rise to an unbiased estimator of the likelihood. 
We use systematic resampling. 
In this resampling scheme, we sample an initial $U_{1} \sim \text{Uniform}(0, 1/K)$ and set $U_{i} = U_{1} + (i-1)/K$. 
We then sample the particles for which the cumulative weights are between $U_{i-1}$ and $U_{i}$. 
Systematic resampling empirically yields lower-variance estimates of the log-likelihood, slowing path degeneracy (\cite{chopin}).

\subsubsection{Adaptive resampling}

In Algorithm \ref{alg:smc} resampling is performed in every step. 
However, we know that this resampling step introduces noise and reduces our number of useful particles. 
One way to reduce this noise is to not resample in every step. 
Instead, we only resample if the number of useful particles, denoted by the effective sample size (ESS), falls below a threshold, conventionally set at $K/2$. 
ESS of normalised weights $W^{1:K}$ is given by:
\begin{equation}
\text{ESS}(W^{1:K}) = \frac{1}{\sum_{k=1}^{K} (W^{k})^{2}} 
\end{equation}
Its interpretation as the number of useful particles is because if one particle has all the weight, then $\text{ESS}=1$, and if all particles have equal weight, then $\text{ESS}=K$.

\subsubsection{Backward simulation}

Running the PMMH algorithm only requires unbiased likelihood estimators, but we are also interested in the realisations of the birth rate trajectories. 
These can be extracted from the SMC algorithm naively by tracking the history of one particle sampled with probability proportional to its weight on the last day. 
However, this is vulnerable to path degeneracy, as particles with high weight on the last day are often descendants of the same particles on early days. 
Therefore, the last strategy we employ to reduce the variance introduced by resampling is a method called backward simulation (\cite{godsill2004}). 

Backward simulation is run backwards in time after the SMC algorithm. 
It re-weights samples to smooth out trajectories such that the samples of the birth rate and prevalence on each day $n \leq N$ are conditional on observations from all days $1, \ldots, N$ rather than only historical days $1, \ldots, n$. 
That is, rather than choosing a sample trajectory $\beta_{1:N}^{k}$ according to the weights on day $N$, we use backward simulation to sample a trajectory. We get this trajectory using the following steps:

\begin{enumerate}
\item Run the SMC algorithm forward-in-time, storing all particles and weights in each generation, even those culled by resampling.
\item Set $j_{N}=k$ with probability $w_{N}^{k} / \sum_{l} w_{N}^{l}$.
\item For $n = N-1, \ldots, 1$, compute the smoothing weights
\begin{equation*}
w_{n \mid N}^{k} = \frac{w_{n}^{k} p(\beta_{n+1}^{j_{n+1}}, x_{n+1}^{j_{n+1}} \mid \beta_{n}^{k}, x_{n}^{k})}{\sum_{l} w_{n}^{l} p(\beta_{n+1}^{j_{n+1}}, x_{n+1}^{j_{n+1}} \mid \beta_{n}^{l}, x_{n}^{l})}.
\end{equation*}
\item Set $j_{n}=k$ with probability $w_{n \mid N}^{k}$.
\end{enumerate}
\medskip

Then our sample trajectory is $\{\beta_{n}^{j_{n}}\}$ for all $n = 1, \ldots, N$. 
This path should be representative across the whole time series, rather than only a good fit at the end.

Empirical results showing the improvements from implementing systematic resampling, adaptive resampling and backward simulation can be seen in Appendix \ref{appendix:pathdegen}.

\subsection{Choosing the number of particles} \label{subsec:nopt}

Analogously to move size in the Metropolis--Hastings algorithm, too few particles will yield sticky chains, and too many particles is computationally inefficient.
To choose an optimal number of particles, we use the suggested guidance from \textcite{pitt2012} which proposes the following steps:

\begin{enumerate}
\item Run a short MCMC scheme with a large number of particles $K$ to determine an approximate value for the posterior mean of $\theta$, denoted $\Bar{\theta}$.
\item Run the SMC algorithm for several (say $R=100$) independent runs for a fixed starting value of particles $K_{s}$ and obtain an estimator of the likelihood $\hat{p}_{K_{s}}^{i}(y \mid \Bar{\theta})$, $i = 1, \ldots, R$ for each.
\item Record the variance of the log of the likelihood estimator as
\begin{equation*}
\hat{\sigma}^{2}(\Bar{\theta}, K_{s}) = \frac{1}{R} \sum_{i=1}^{R} \left(\log \hat{p}_{K_{s}}^{i}(y \mid \Bar{\theta}) \right)^{2} - \left( \frac{1}{R} \sum_{i=1}^{R} \log \hat{p}_{K_{s}}^{i}(y \mid \Bar{\theta}) \right)^{2}.
\end{equation*}
\item Choose the optimal number of particles $K_{opt}$ as
\begin{equation*}
K_{opt} = K_{s} \times \frac{\hat{\sigma}^{2}(\Bar{\theta}, K_{s})}{0.92^{2}}.
\end{equation*}
\end{enumerate}

0.92 is the number calculated in \textcite{pitt2012} which minimises the computing time of the pseudo-marginal MCMC algorithm. 
In practice we found that this method can be volatile, because step 1 does not always manage to find a good value for $\Bar{\theta}$. 
We instead run this scheme three times and choose the maximum number of particles.

\FloatBarrier

\section{Simulation study} \label{sec:simulation}

\FloatBarrier

All trace plots and posterior density plots for parameters $\sigma$ and $X_{0}$ for this section and the next section can be found in Appendix \ref{appendix:trace}. 
We used the scheme described in Section \ref{subsec:nopt} to choose the number of particles, subject to a cap of $25,000$ particles. 
This ceiling on the number of particles imposes an upper limit on computational cost. 
This is required due to a 48-hour runtime limit on our HPC cluster. 
We also set a minimum number of particles of 1,000 as a precautionary measure against choosing too few particles.
All chains chose the floor of 1,000, so the maximum run time was 31 minutes.

\subsection{Priors} \label{subsec:priors}

The prior used for $\beta_{1}$, the birth rate on the first day of the epidemic, is exponential with rate $(2\gamma)^{-1}$. 
This has mean $2\gamma$ to reflect a belief that the epidemic will be growing initially. 
The prior for $\beta_{n} \mid \beta_{n-1}, \sigma$ is a $\text{Folded-Normal}(\beta_{n-1}, \sigma^{2})$ for $n = 2, \ldots, N$. 
The folded normal distribution is the distribution of the absolute value of a random variable with a normal distribution (\cite{leone1961}).

The prior used for $\sigma$ is exponential with rate 10. 
This has a mean of 0.1 as we expect small changes in the birth rate between days.

The prior used for the reporting probability $\rho$ is Uniform$(0,1)$. 
This allows us to suppose that we have no notion of how much of the epidemic that we are observing.

The prior used for $X_{0}$, the number of cases on day 0, is a negative binomial distribution with mean 5 and variance 50. 
This reflects a prior belief that the prevalence on day 0 should be small, but there is uncertainty around this.
Using the method of moments, this gives a number of successes $r=0.56$ and a success probability $p=0.1$, so $X_{0} \sim \text{Negative-Binomial}(0.56, 0.1)$.

\FloatBarrier

\subsection{Peaked reproduction number}

\FloatBarrier

\begin{figure}[ht]
\centering
\includegraphics[width=\textwidth]{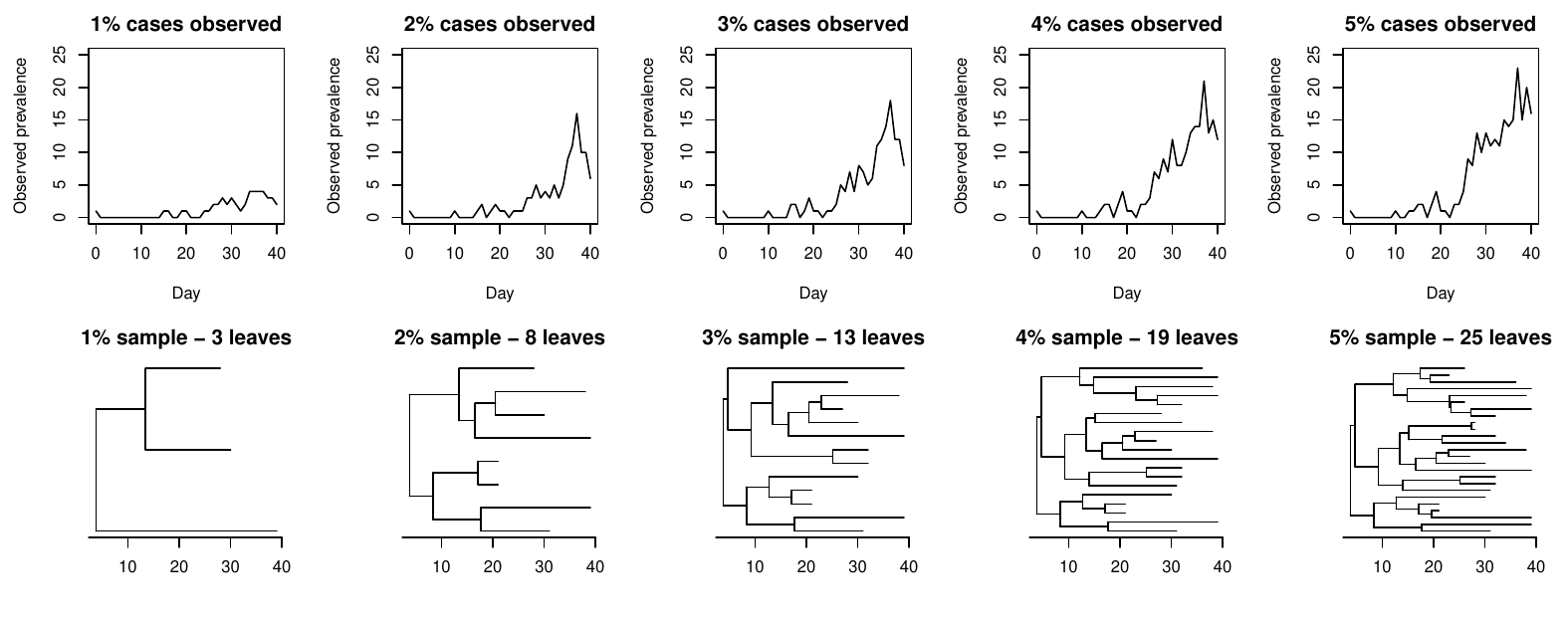}
\caption{Observed prevalences and sampled phylogenies of a simulated epidemic with a peaked birth rate.}
\label{fig:data_peak}
\end{figure}

For our simulation example, we consider the realistic scenario of a time-varying reproduction number that is increasing smoothly and then decreasing smoothly. 
We have simulated a 40-day epidemic with a birth rate increasing linearly from 0.1 to 0.3 until midway through the epidemic and then decreasing linearly from 0.3 to 0.1, and a death rate of $\gamma=0.1$s. 
This is equivalent to starting with $R_{t=0}=1$, increasing linearly to $R_{20}=3$ and then decreasing linearly to $R_{40}=1$. 
The observed prevalences and sampled phylogenies are shown in Figure \ref{fig:data_peak}.

\begin{figure}[ht]
\centering
\includegraphics[width=\textwidth]{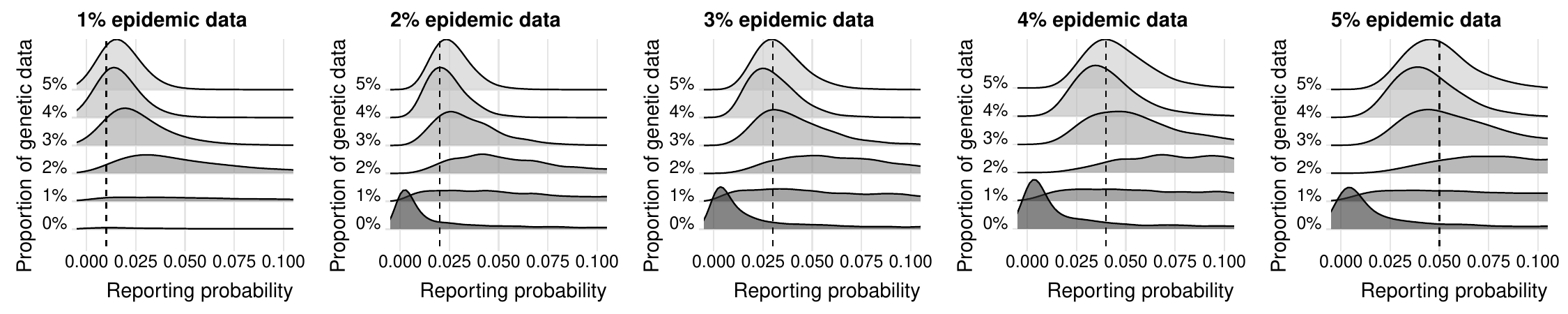}
\caption{Posterior density of the reporting probability for a simulated epidemic with a peaked birth rate. The dashed line represents the true reporting probability.}
\label{fig:pobs_peak}
\end{figure}

\begin{figure}[ht]
\centering
\includegraphics[width=\textwidth]{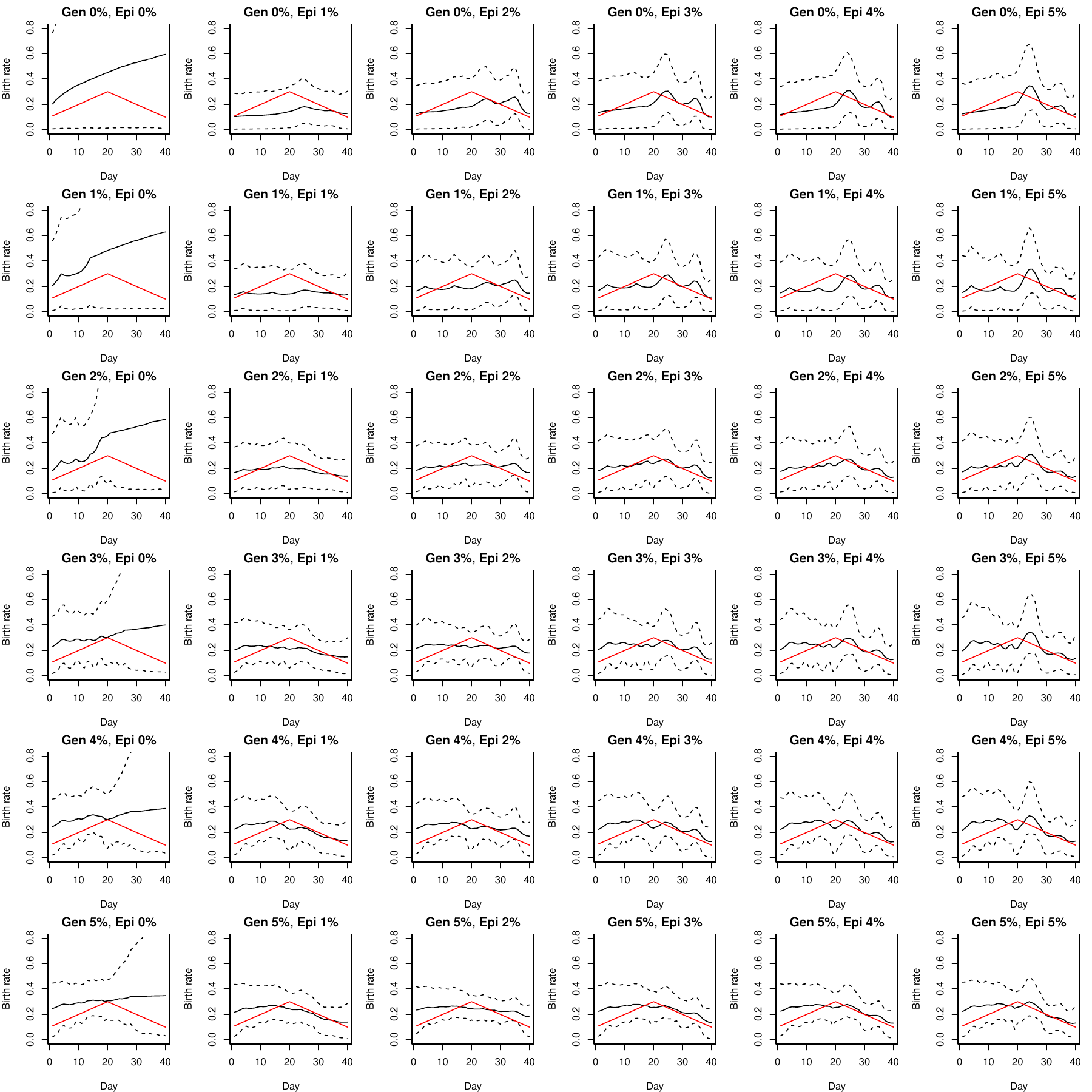}
\caption{
Inferred birth rates for a simulated epidemic with peaked birth rate. 
The solid black lines show the posterior mean, the dashed lines show the posterior 95\% credible interval, and the true trajectory is shown in red.
}
\label{fig:br_peak}
\end{figure}

For each sampling proportion from 0\% to 5\%, we have run the PMMH algorithm. 
All chains have been run for $100,000$ iterations with initial values of $\sigma_{0}=0.05$, $\rho_{0}=0.03$ and $X_{0}=1$. 
Posterior density plots for $\rho$ are shown in Figure \ref{fig:pobs_peak}. 
Inference plots for $\beta_{t}$ are shown in Figure \ref{fig:br_peak}.

It is clear from the posterior density plots of $\rho$ in Figure \ref{fig:pobs_peak} that the inclusion of genetic data improves the inference of the reporting probability compared to epidemiological data alone. 
In fact, more genetic data seems to improve inference of the reporting probability. 
For each level of epidemic data, the 3\% tree with 13 leaves starts to show a marked improvement in inference of $\rho$. 
It is worth noting that 13 leaves is a small amount of genetic data. 
Initially that genetic data would improve inference of the reporting probability may seem counter-intuitive as the phylogeny does not contain any information about the reporting probability directly. 
However, we hypothesise that this improvement is due to the genetic data aiding the inference of the effective population size and therefore the ratio between the birth rate and the true prevalence. 
For all scenarios with data, the posterior mean for $\rho$ is closer to the true value than its prior mean of 0.5, and the shape of the distribution is clearly non-uniform, so all posteriors show learning from the data.

Figure \ref{fig:br_peak} shows reasonably good inference, considering the small amounts of data. 
For all levels of data, the trajectory is inferred after the peak well, but poorly before the peak. 
This is because in a peaked scenario, most genetic data is concentrated in the middle of the epidemic and most prevalence data is at the end of the epidemic, so it can be difficult to infer further back in time than midway. 

Overall, the inclusion of genetic data seems to improve the inference of $R_{t}$ and reduce the credible intervals, without excluding the truth. 
When there is 1\% or more of epidemic data, the mean of the root mean square errors (RMSE) when there is no genetic data is 0.0585 and when there is 5\% genetic data the mean RMSE is 0.0570. 
When there is 1\% or more of epidemic data, the mean credible interval width when there is no genetic data is 0.36 and when there is 5\% genetic data is 0.26.

\FloatBarrier

\section{Case study: HIV-1 in North Carolina, USA}

\FloatBarrier

\subsection{PMMH for HIV-1}

\begin{figure}[ht]
\centering
\includegraphics[width=0.95\linewidth]{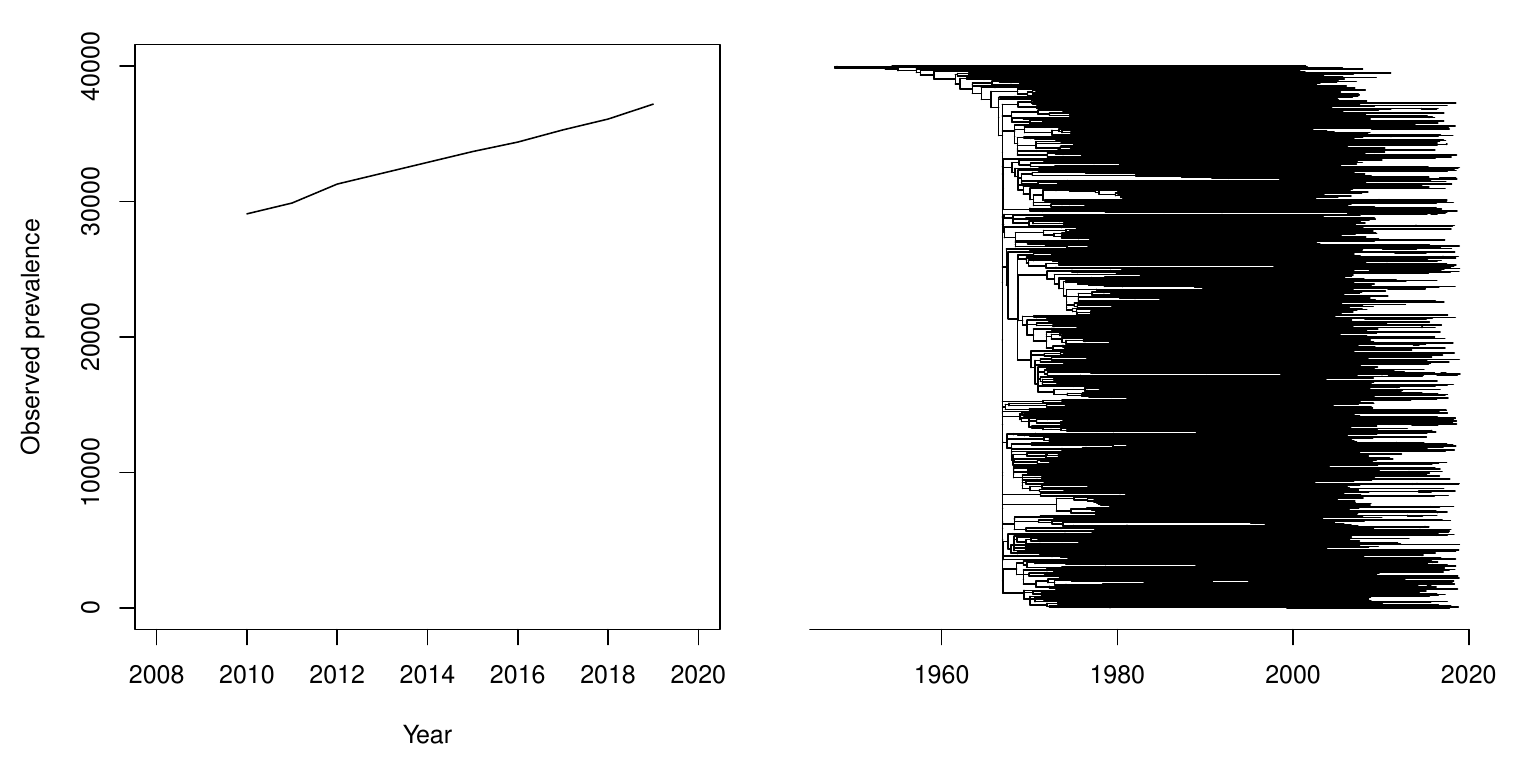}
\caption{Observed prevalence of HIV-1 (left) and phylogenetic tree of HIV-1 (right).}
\label{fig:hiv_data}
\end{figure}

Having verified the method in Section \ref{sec:simulation}, we now consider a real data set. 
This data is of HIV-1 in North Carolina, USA. 
Sequences were sampled from 1997 to 2019 and the US Centers for Disease Control had estimates of the prevalence of HIV in North Carolina from 2010 to 2019. 
This data is from \textcite{dennis2021} and \textcite{didelot2023} and more details can be found there. 
The prevalence and the phylogenetic tree are shown in Figure \ref{fig:hiv_data}.

The phylogeny has 1850 leaves.
The first sequence was taken on 1st May 1997 and the last was on 15th January 2019.
The most recent common ancestor of the sample is on 27th November 1947.

The first reported cases of HIV-1 were in 1981, so we have run the PMMH algorithm from 1980 to 2019. 
The prior we use for the smoothness $\sigma$ is exponential with rate 10.
The prior for the reporting probability $\rho$ is Uniform$(0,1)$.
Lastly, we use a negative binomial prior for the prevalence in 1980 with $r=514.14$ and $p=0.117$.

After running several short MCMC runs on the data, we found that number of particles needed to achieve good mixing with the HIV-1 data was much larger than simulations, so we imposed a cap on the number of particles of 50,000. 
Due to a 48-hour run time limit on our HPC cluster, this meant that an MCMC chain could only run for around 25,000 to 30,000 iterations. 
As such, rather than running one long chain, we have run 4 short chains and then pooled the MCMC samples for the analysis. 
Initial values were chosen carefully from the several short MCMC runs in order to minimise burn-in.

\begin{figure}[ht]
\centering
\includegraphics[width=\textwidth]{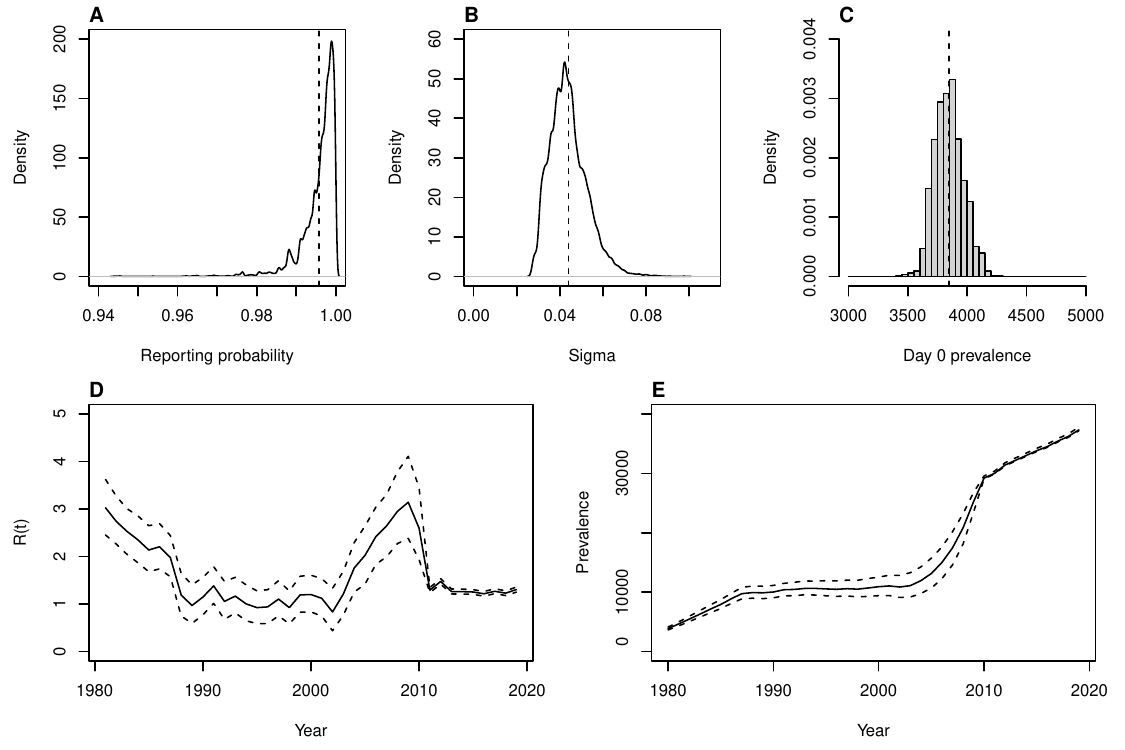}
\caption{
A: The posterior density of the reporting probability $\rho$. 
B: The posterior density of $\sigma$. 
C: The posterior density of the initial prevalence $X_{0}$. 
A-C: The dashed vertical line represents the posterior mean. 
D: The inferred reproduction number from the HIV-1 data set. 
E: The inferred prevalence for the HIV-1 data set. 
D-E: The solid line is the posterior mean and the dashed lines represent the posterior 95\% credible interval.}
\label{fig:hiv_inference}
\end{figure}

Initial values used were $\rho_{0}=0.99$, $\sigma_{0}=0.05$ and $X_{0}=3750$. 
A large $\rho_{0}$ is because we expect the reporting probability of HIV-1 to be high as it is a high profile disease that is regularly tested for. 
A small $\sigma_{0}$ is because we expect small changes in the reproduction number between years. 
The initial value $X_{0}=3750$ initialised the chains at there being 3,750 cases of HIV-1 in 1980. 
The death rate used was $\gamma=0.1$ to reflect a removal time of 10 years. 
The optimum number of particles chosen was 50,000 for all chains and the chains ran for 48 hours. 
On average the chains ran for 26,219 iterations in the 48 hours. 
The target acceptance rate was 10\% and the average acceptance rate was 7.2\%. 
The results of this inference can be seen in Figure \ref{fig:hiv_inference}.

The posterior densities of the parameters $\rho$, $\sigma$ and $X_{0}$ all look considerably different to the priors, indicating that we have learned from the data. 
The posterior mean of the reporting probability is 99.6\%. 
This is surprisingly high as it is estimated that $87\%$ of people with HIV in the USA know that they have HIV (\cite{hiv}). 
The posterior mean for $X_{0}$ is 3,846, which is close to the suspected initial value of 3,750. 
This suggests there were approximately 3,846 cases of HIV-1 in 1980. 
This is indeed low, although it should be noted that we only have data from North Carolina, so cannot generalise to the whole USA. 
The posterior mean for $\sigma$ is 0.044, which is lower than the prior mean of 0.1. 
This indicates that $R_{t}$ is generally changing by at most $\pm 0.88$ (i.e. $\pm$ 2 standard deviations) per year.

Inference of the latent prevalence spans 1980 to 2019. 
We have only phylogenetic data from 1980 to 2009 and both phylogenetic and prevalence data from 2010 to 2019. 
The inferred prevalence has an S-shape to it whereby cases increased slowly to around 1987, level out until around 2004, rapidly increase again until around 2010, and then increase slowly again to 2019. 
This may be because antiretroviral therapy was introduced in 1995, so more people were living with HIV-1 rather than dying from it, leading to a sharper increase in prevalence in the 2000s. 
A similar S-shaped prevalence trajectory can be seen in the National Institute on Drug Abuse's HIV/AID Research Report (\cite{nida}).

Inference of the reproduction number spans from 1981 to 2019. 
We have only phylogenetic data from 1981 to 2009 and both phylogenetic and prevalence data from 2010 to 2019. 
The inclusion of the epidemiological data to the phylogenetic data can be seen in the width of the credible interval. 
The mean width of the credible intervals from 1981 to 2009 is 0.95 and from 2010 to 2019 is 0.25. 
The combination of phylogenentic information and prevalence seems to provide a lot more certainty in the inference of the reproduction number compared with phylogenetic data alone. 
This is consistent with what we saw in the simulations, that the credible intervals using our method can be wide unless incorporating both sources of data. 
The posterior mean in Figure \ref{fig:hiv_inference}D suggests that the effective reproduction number for HIV-1 in North Carolina was above 1 from 1981 to 1987, around 1 from 1988 to 2003, and then above 1 from 2004 to 2019. 
The posterior 2.5\% quantile is above 1 from 1981 to 1987 and from 2004 to 2019, so we can be confident that HIV-1 was spreading during these times. 
Despite the width of the credible intervals, the phylogenetic data allows us to infer $R_{t}$ with reasonable confidence much further back in time than the prevalence data alone could allow.

\FloatBarrier
\subsection{Comparison to existing methods for HIV-1}
\FloatBarrier

\begin{figure}[ht]
\centering
\includegraphics[width=\textwidth]{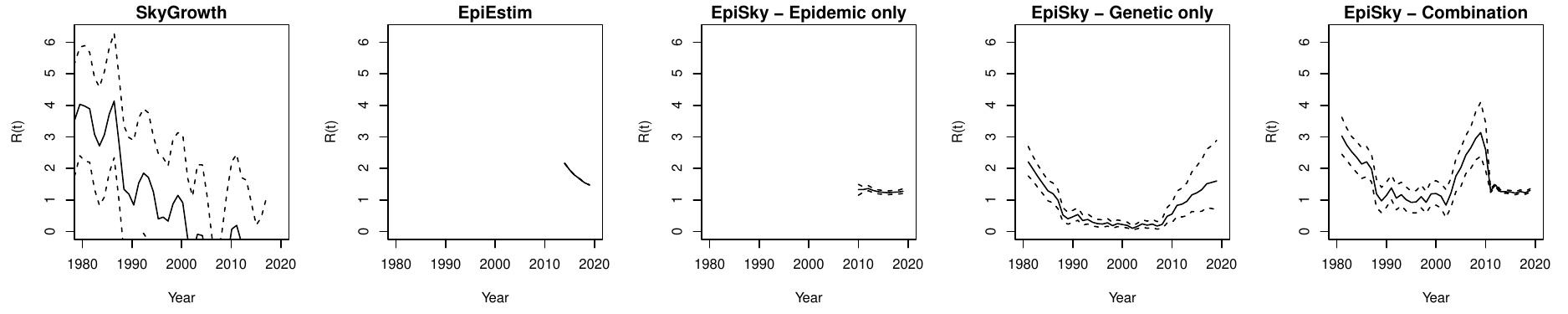}
\caption{Comparison of $R_{t}$ inferred by EpiEstim, SkyGrowth and our PMMH approach with only epidemic data, only genetic data and combination data. The solid lines represent posterior means and the ribbons represent 95\% credible intervals.}
\label{fig:hiv_comparison}
\end{figure}

We will compare our results to EpiEstim, an R package used to estimate time-varying reproduction numbers from epidemic curves (\cite{cori2013}), and to SkyGrowth, a GitHub repository for phylodynamic inference in R using time-scaled phylogenies (\cite{volz2018}). 
This comparison is shown in Figure \ref{fig:hiv_comparison}.

The serial interval is the time from symptom onset in a primary case to symptom onset in a secondary case.
To use EpiEstim, we are required to have estimates for the mean and standard deviation of the serial interval. 
As in \textcite{safina2022}, we will use the estimates from \textcite{hollingsworth2008} to construct the distribution of serial intervals. 
The most recent common ancestor in the phylogeny is in 1947, so we ran the MCMC within SkyGrowth ran for 100,000 iterations using 72 time points to give one time point per year from 1947 to 2019.

In this case, EpiEstim and our PMMH approach are not in good agreement with each other. 
This is likely due to the difficulty of deducing the serial interval for HIV-1, which is the basis of the analysis in EpiEstim. 
Our approach does not require knowledge of the serial interval, so it is more flexible. 

There are many years where the inference from SkyGrowth and PMMH are compatible with each other, although the width of the credible intervals for SkyGrowth is much larger than for PMMH. 
The broad pattern SkyGrowth is showing is that $R_{t}$ is decreasing from 1980 to 2018. 
SkyGrowth does not see the increase in $R_{t}$ from 2002 that PMMH using combination data and only genetic data picks up. 
SkyGrowth does not estimate $R(t)$ directly. 
Instead it estimates the growth rate of the population $r(t)$ and used the relationship $R(t)=r(t)/(\gamma+1)$ to relate the growth rate to the reproduction number. 
When the growth rate is smaller than the death rate (i.e. the population is shrinking faster than would be expected even from a birth rate of 0), this relationship can give values below 0. 
As this is not biologically meaningful, we have cropped the y-axis below at $R(t)=0$. 
However, this is a limitation in SkyGrowth that is not present when using our PMMH approach.

\FloatBarrier

\section{Case study: Tuberculosis in Argentina}

\FloatBarrier
\subsection{PMMH for TB}

\begin{figure}[ht]
\centering
\includegraphics[width=0.95\linewidth]{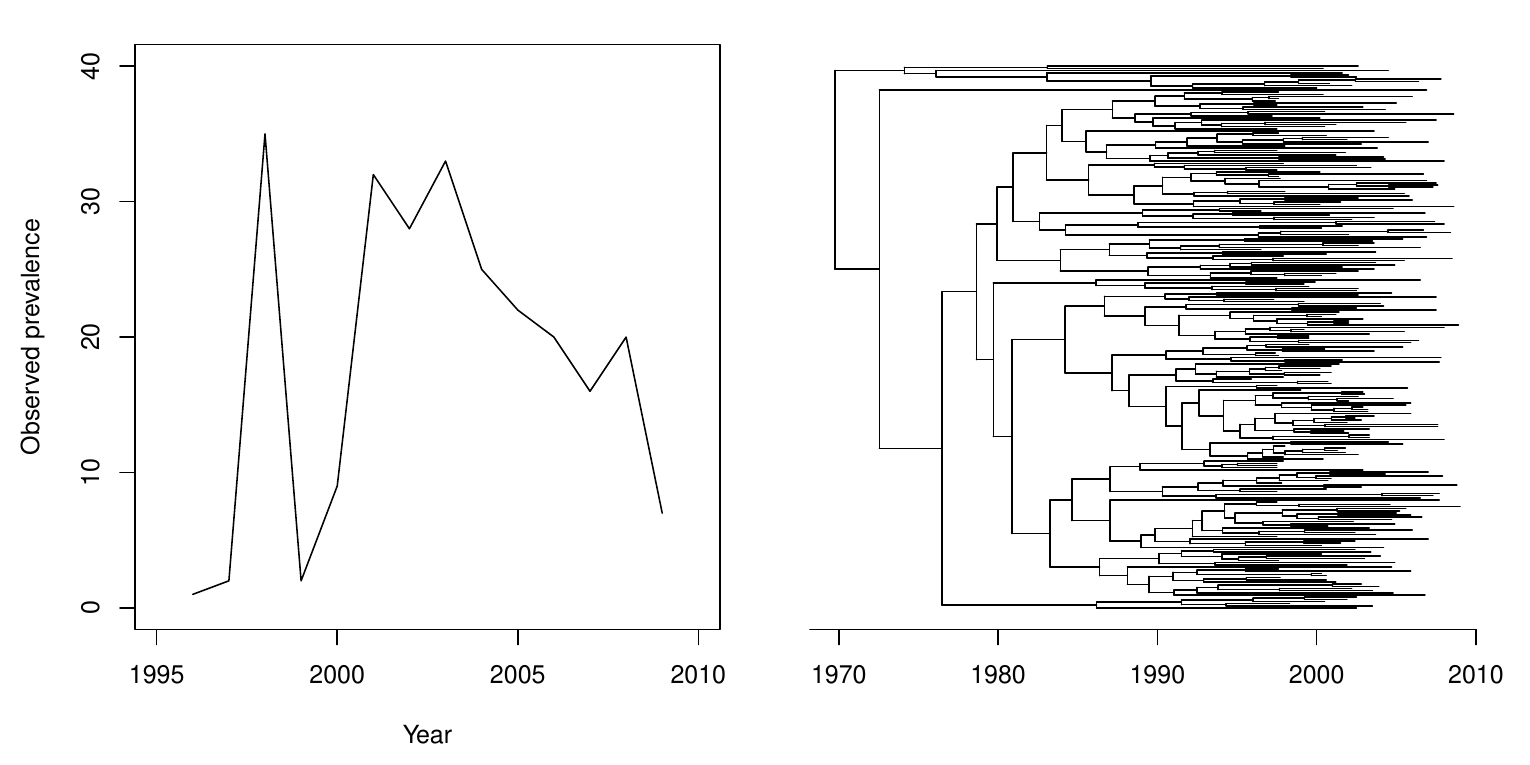}
\caption{Observed prevalence of TB (left) and phylogenetic tree of TB (right).}
\label{fig:tb_data}
\end{figure}

In this section we will consider a real data set of drug-resistant tuberculosis in Buenos Aires, Argentina collected between 1996 and 2009. 
This data comes from \textcite{tb} and more details of the phylogenetic reconstruction can be found there. 
We will use the number of isolates collected as the observed prevalence data. 
The prevalence and the phylogenetic tree are shown in Figure \ref{fig:tb_data}.

The prior we use for the smoothness $\sigma$ is exponential with rate 10.
The prior for the reporting probability $\rho$ is Uniform$(0,1)$.
Lastly, we use a negative binomial prior for the prevalence in 1980 with $r=1$ and $p=0.125$.

\begin{figure}[ht]
\centering
\includegraphics[width=\textwidth]{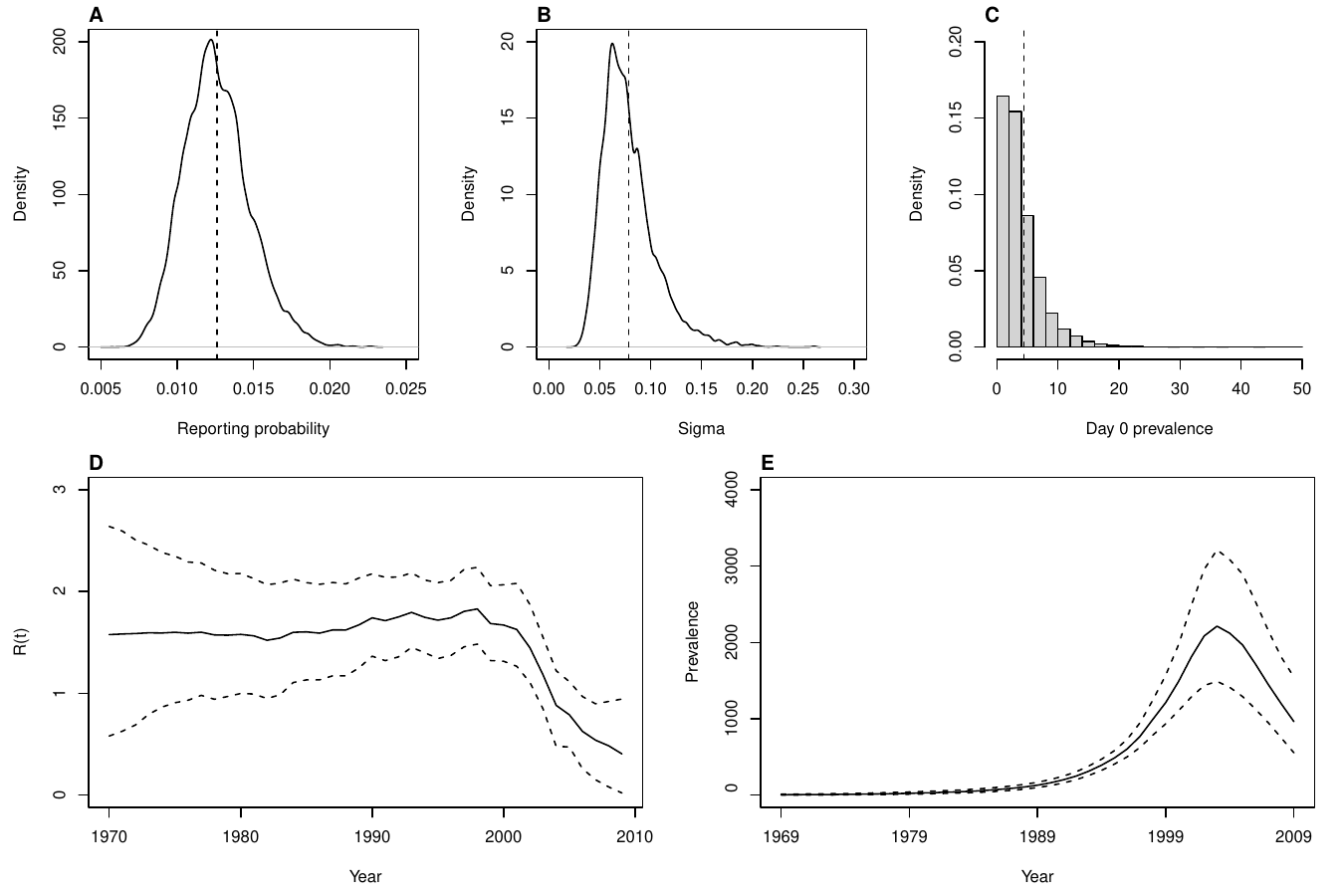}
\caption{
A: The posterior density of the reporting probability $\rho$. 
B: The posterior density of $\sigma$. 
C: The posterior density of the initial prevalence $X_{0}$. 
A-C: The dashed vertical line represents the posterior mean. 
D: The inferred reproduction number from the TB data set. 
E: The inferred prevalence for the HIV-1 data set. 
D-E: The solid line is the posterior mean and the dashed lines represent the posterior 95\% credible interval.}
\label{fig:tb_inference}
\end{figure}

We have imposed a cap of 50,000 on the number of particles in order to cap the run time to around 1 day. 
We have also imposed a minimum number of particles of 10,000 to ensure mixing.
The PMMH algorithm was run for 100,000 iterations and the initial values used were $\rho_{0}=0.05$, $\sigma_{0}=0.1$ and $X_{0}=5$. 
The death rate used was $\gamma=1/3$ to reflect a removal time of 3 years. 
The number of particles chosen was 10,000 and the chain took 11 hours and 31 minutes to run. 
The target acceptance rate was 10\% and the actual acceptance rate was 10.3\%. 
The results of this inference can be seen in Figure \ref{fig:tb_inference}.

The posterior densities of the parameters $\rho$, $\sigma$ and $X_{0}$ all look considerably different to the priors, indicating that we have learned from the data. 
The posterior mean of the reporting probability is 1.3\%, so our analysis suggests that only 1.3\% of the true number of TB cases are in our sample. 
The posterior mean for $X_{0}$ is 4.4, which is close to the suspected initial value of 5. 
This indicates that our sequences date back to a small number of lineages from the 1970s. 
The posterior mean for $\sigma$ is 0.08, which is close to the prior mean and initial value of 0.1.
This indicates that $R_{t}$ is generally changing by at most $\pm 0.47$ (i.e. $\pm$ 2 standard deviations) per year. 

Inference of the reproduction number and latent prevalence spans 1970 to 2009. 
We only have phylogenetic data from 1970 to 1995 and both phylogenetic and prevalence data from 1996 to 2009. 
The posterior mean in Figure \ref{fig:tb_inference}D suggests that the effective reproduction number for TB in Buenos Aires was above 1 from 1970 to 2003 and then below 1 from 2004 to 2009. 
However, we can only be reasonably confident that $R_{t}$ was above 1 in 1981 and from 1984 to 2002, as this is when the 2.5\% lower credible interval is above 1. 
The prevalence of TB seems to be increasing until 2003 and then decreasing thereafter, which is consistent with $R_{t}$ dropping and staying below 1 from 2004.

\FloatBarrier
\subsection{Comparison to existing methods for TB}
\FloatBarrier

\begin{figure}[ht]
\centering
\includegraphics[width=\textwidth]{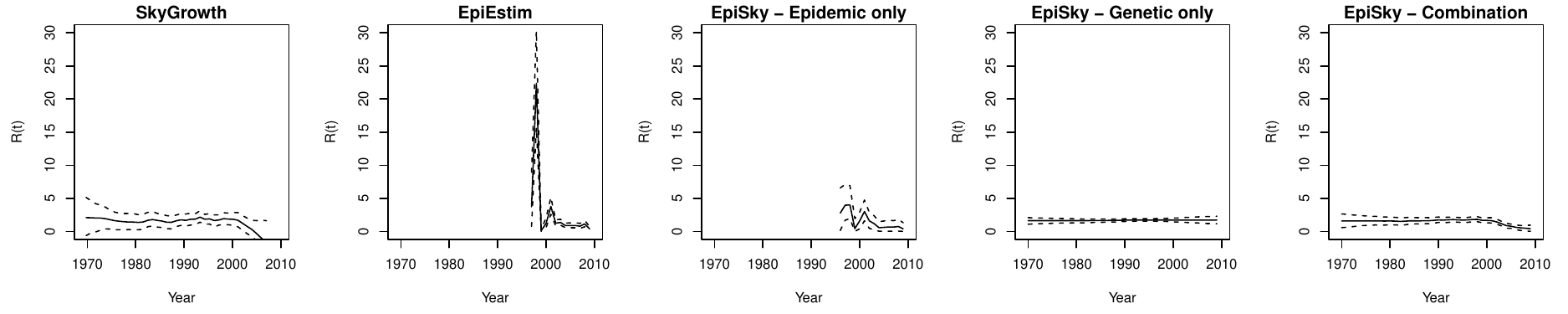}
\caption{Comparison of $R_{t}$ inferred by EpiEstim, SkyGrowth and our PMMH approach with only epidemic data, only genetic data and combination data. The solid lines represent posterior means and the ribbons represent 95\% credible intervals.}
\label{fig:tb_comparison}
\end{figure}

As before, we compare our results to EpiEstim and SkyGrowth, see Figure \ref{fig:tb_comparison}. 
For EpiEstim, we use a mean serial interval of 2, as estimated in \textcite{ma2020} to be the serial interval in Brazil from March 2008 to June 2012. 
We use a standard deviation in the serial interval of 1. 
The MCMC within SkyGrowth ran for 100,000 iterations using 40 time points.

From 1999 to 2009, EpiEstim and the PMMH using only prevalence data are in good agreement with one another. 
However, in 1998 the observed prevalence spiked from 2 in 1997 to 35 in 1998, so EpiEstim interprets this as a very large $R_{t}$ of approximately 22 in 1998. 
This is highly likely to be an artificial feature of the noisy prevalence data, so the PMMH approach appears to be more robust in this case. 
In comparison, the estimates using the PMMH algorithm with combination data is much smoother and the credible intervals are tighter, showing our approach is beneficial methodologically and that the addition of genetic data helps with epidemiological inference.

As with EpiEstim, there are many years where the inference from SkyGrowth and PMMH are compatible with each other, although the width of the credible intervals for SkyGrowth is much larger than for PMMH. 
When we compare SkyGrowth to PMMH using only the genetic data, they are in generally good agreement up to 2000, where SkyGrowth detects the drop in $R_{t}$ that PMMH only sees with combination data. 
However, the credible intervals given by SkyGrowth contain negative values from 1970 to 1972 and 2004 to 2009, and indeed, the mean $R_{t}$ value inferred is negative from 2006 to 2009.

Overall, our estimates are consistent with current approaches when appropriate and more robust when not.

\FloatBarrier

\section{Discussion}

\FloatBarrier

\subsection{Summary of findings}

In this paper, we have introduced using particle MCMC methods, specifically the PMMH algorithm, to infer the reproduction number of an epidemic over time using a dated phylogeny and partially observed prevalence data. 

In many simulated scenarios, the combination of data sources offers improvement in the inference and certainty of the inference in the reproduction number compared with one source or the other. 
In particular, the introduction of a small amount of epidemiological data to the genetic data drastically improves the inference. 
This is useful in practice as we were primarily considering the setting with poor epidemiological data.
It should be feasible in general to construct a decent phylogenetic tree, because inference on phylogenetic trees is robust to taking subtrees.
We also found in simulations that the introduction of genetic data to the epidemiological data vastly improves the inference of the reporting probability $\rho$.

We have then performed inference on real data sets. 
Inference in a real scenario is consistent with simulation results, in that inference of the reporting probability $\rho$ is improved with combination data, as is certainty of $R_{t}$ inference. 
We were able to compare the inferred latent prevalence to other sources to confirm that the inference seems to be giving accurate results. 
The real data inference also highlights that combination data allows for inference over a wider period of time than one source of data alone may allow.

\subsection{Strengths and limitations}

Whilst results look very promising, there are limitations to the approach. 
The epidemic model itself is rather simple. 
Compartment models are much more typical in epidemiology, such as susceptible-infectious-removed (SIR) or susceptible-exposed-infectious-recovered (SEIR) models. 
However, the simplicity of the model may not be an issue for inference when considering population-level data. 
The reproduction number is the ratio between $\beta_{t}$ and $\gamma_{t}$ in an SIR model too, the most common epidemiological model, so the current birth-death model may still perform well in cases where an underlying SIR model would be more typically used. 
However, more work would be needed to extend the model to compartment models with more than 3 compartments, such as SEIR models or models with infectious people split by covariates. 
Our method would need to be tested on simulations of other underlying epidemic models and/or real data to examine its robustness to misspecification.

In practice, the dated phylogeny would be estimated using sequencing data, and this paper does not address the uncertainty inherent in the dated phylogeny. 
However, methods such as BEAST and BEAST2 generate posterior phylogenies given sequencing data using MCMC, so we could incorporate this uncertainty by running our method on more than one replicate of the phylogeny output from BEAST/BEAST2. 
We can also see from our simulations that the inference from the genetic trees does not vary much as we change the size of the tree, so we may expect that the additional uncertainty will be fairly small.

Addressing these limitations and thus extending the method to a wider range of applications forms the basis for future work.

\section*{Acknowledgements}

Alicia Gill was supported by EPSRC grant EP/R513374/1. 
Jere Koskela was supported by EPSRC research grant EP/V049208/1. 
Xavier Didelot received funding from the NIHR Health Protection Research Unit in Genomics and Enabling Data. 
Richard G. Everitt was supported by EPSRC grant EP/W006790/1 and NERC grant NE/T00973X/1. 

\addcontentsline{toc}{section}{References}
\printbibliography

\clearpage
\appendix

\FloatBarrier

\section{Appendix - Reducing path degeneracy} \label{appendix:pathdegen}

\FloatBarrier

\begin{figure}[ht]
\centering
\includegraphics[width=\textwidth]{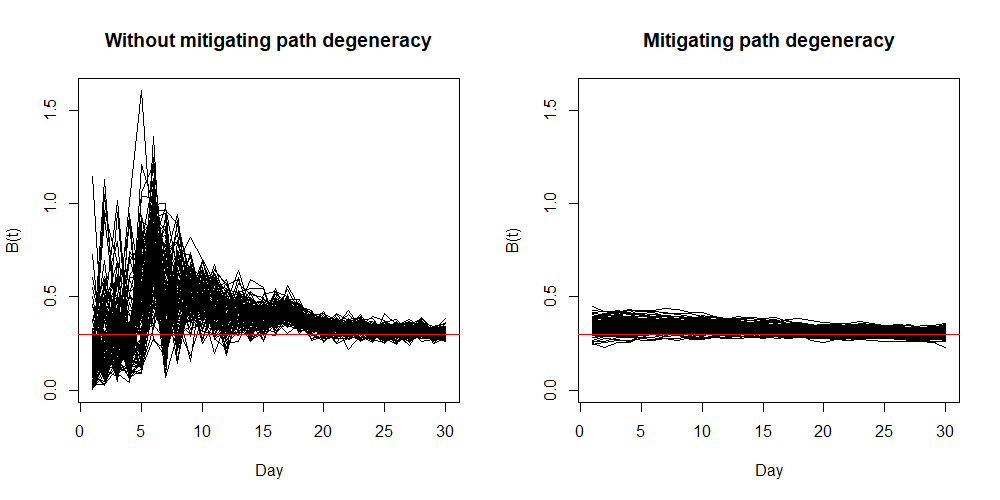}
\caption{Comparison of birth rate inference with and without methods to reduce path degeneracy.}
\label{fig:pathdegen}
\end{figure}

To demonstrate empirically the improvements from implementing systematic resampling, adaptive resampling and backward simulation, we have run some simulations which can be seen in Figure \ref{fig:pathdegen}. 
In these simulations, the true birth rate is constant at 0.3, so $\sigma$ should be close to 0. 
We suppose that 5\% of the true prevalence has been observed, so the true $\rho$ is 0.05. 
We have run the SMC algorithm 100 times with $\sigma = 0.01$, $\rho = 0.05$ and $K=1000$, and for each run output a sample trajectory of the birth rate. 
On the left are the 100 birth rate trajectories with  multinomial resampling, resampling in every step and no backward simulation. 
On the right are the 100 birth rate trajectories with systematic resampling, only resampling if the ESS falls below $K/2$ and backward simulation.

\FloatBarrier

\clearpage
\section{Appendix - Performance diagnostics} \label{appendix:trace}

\FloatBarrier

\begin{figure}[ht]
\centering
\includegraphics[width=\textwidth]{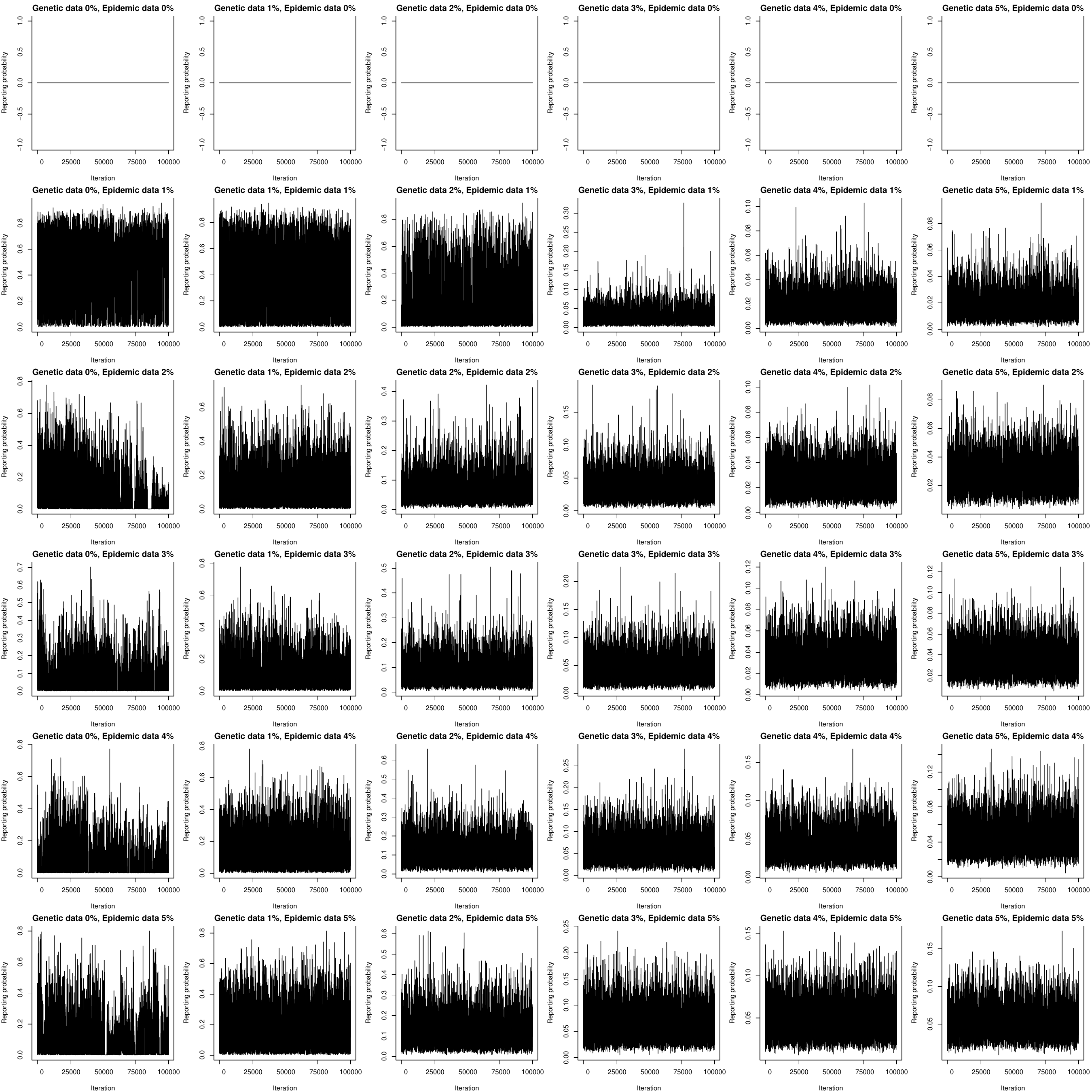}
\caption{Trace plots for $\rho$ for a simulated epidemic with a peaked birth rate.}
\end{figure}\begin{figure}[ht]
\centering
\includegraphics[width=\textwidth]{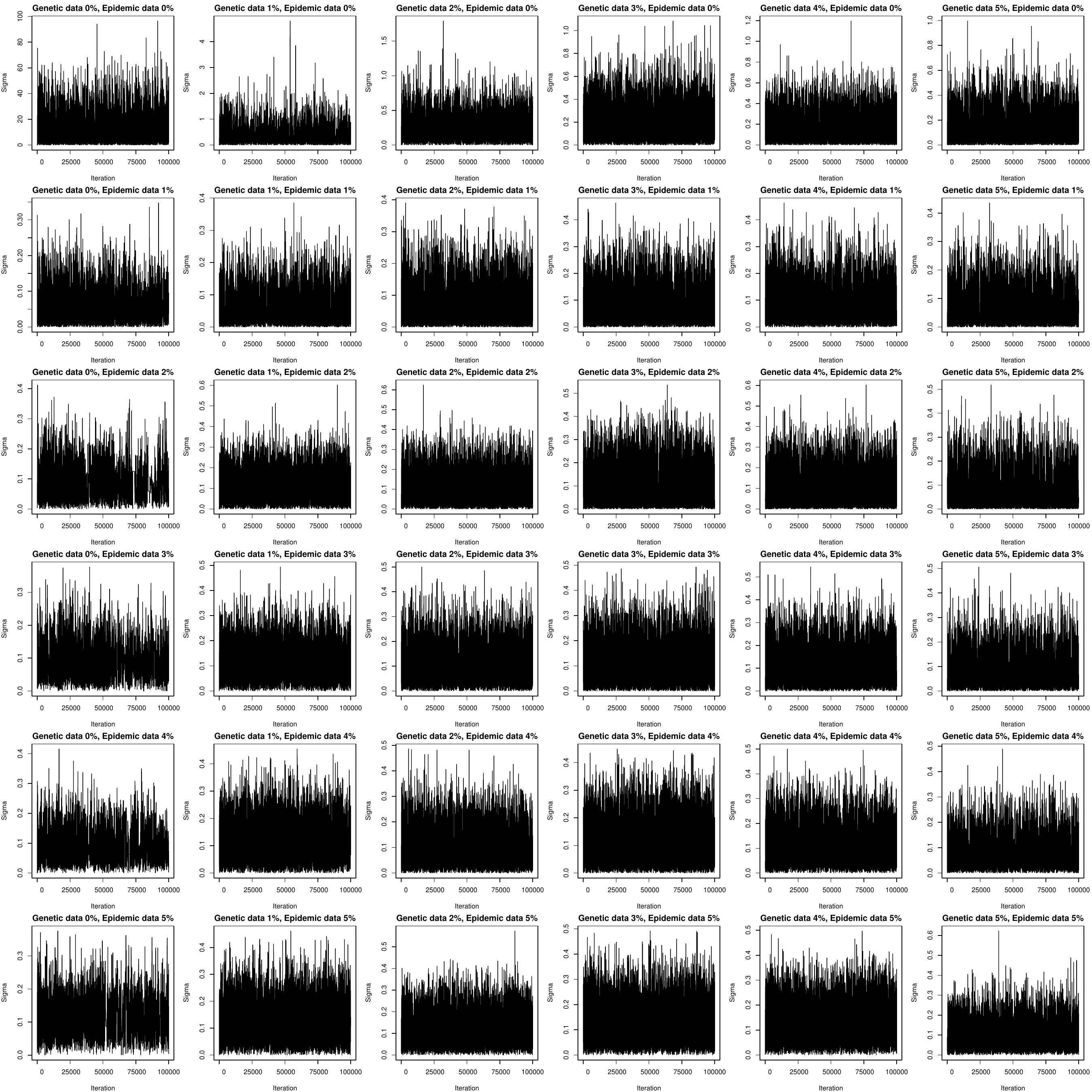}
\caption{Trace plots for $\sigma$ for a simulated epidemic with a peaked birth rate.}
\end{figure}
\begin{figure}[ht]
\centering
\includegraphics[width=\textwidth]{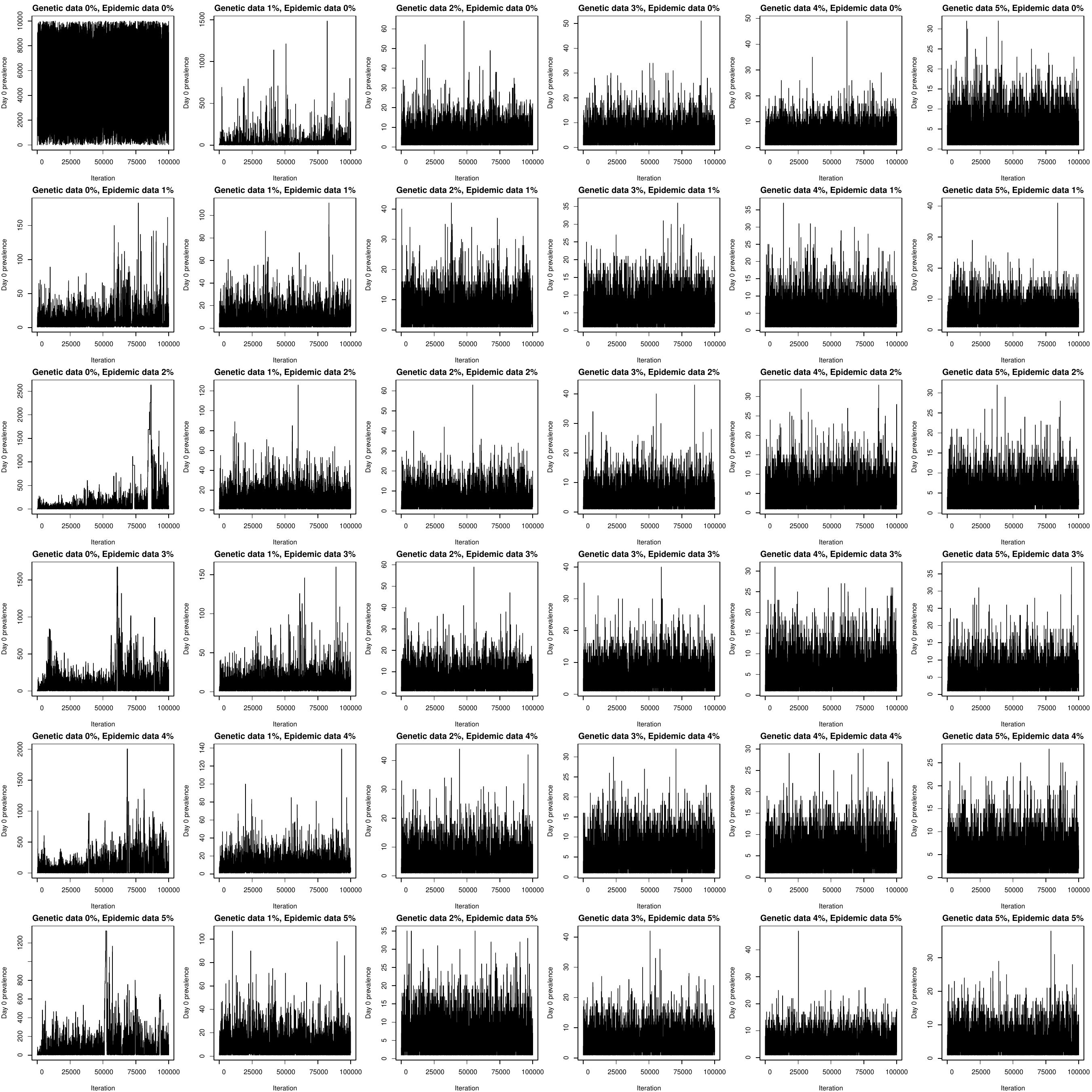}
\caption{Trace plots for $X_{0}$ for a simulated epidemic with a peaked birth rate.}
\end{figure}

\begin{figure}[ht]
\centering
\includegraphics[width=\textwidth]{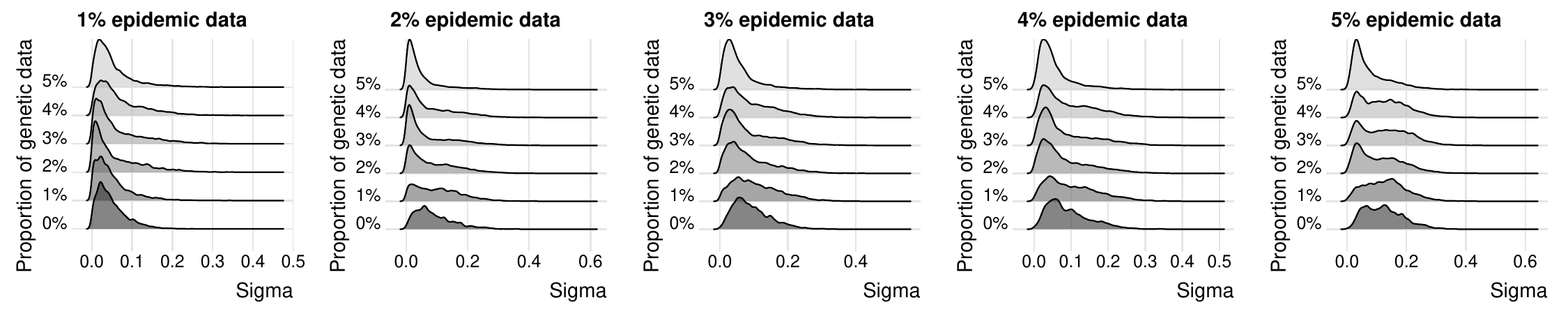}
\caption{Posterior density of $\sigma$ for a simulated epidemic with a peaked birth rate.}
\label{fig:sigma_peak}
\end{figure}
\begin{figure}[ht]
\centering
\includegraphics[width=\textwidth]{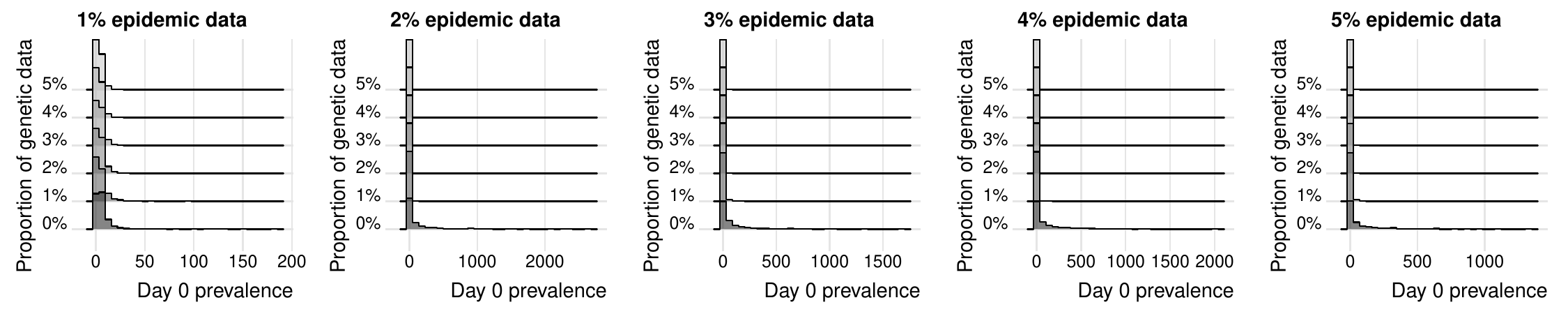}
\caption{Posterior density histogram of the number of cases on day 0 for a simulated epidemic with a peaked birth rate.}
\label{fig:x0_peak}
\end{figure}

\begin{figure}[ht]
\centering
\includegraphics[width=\textwidth]{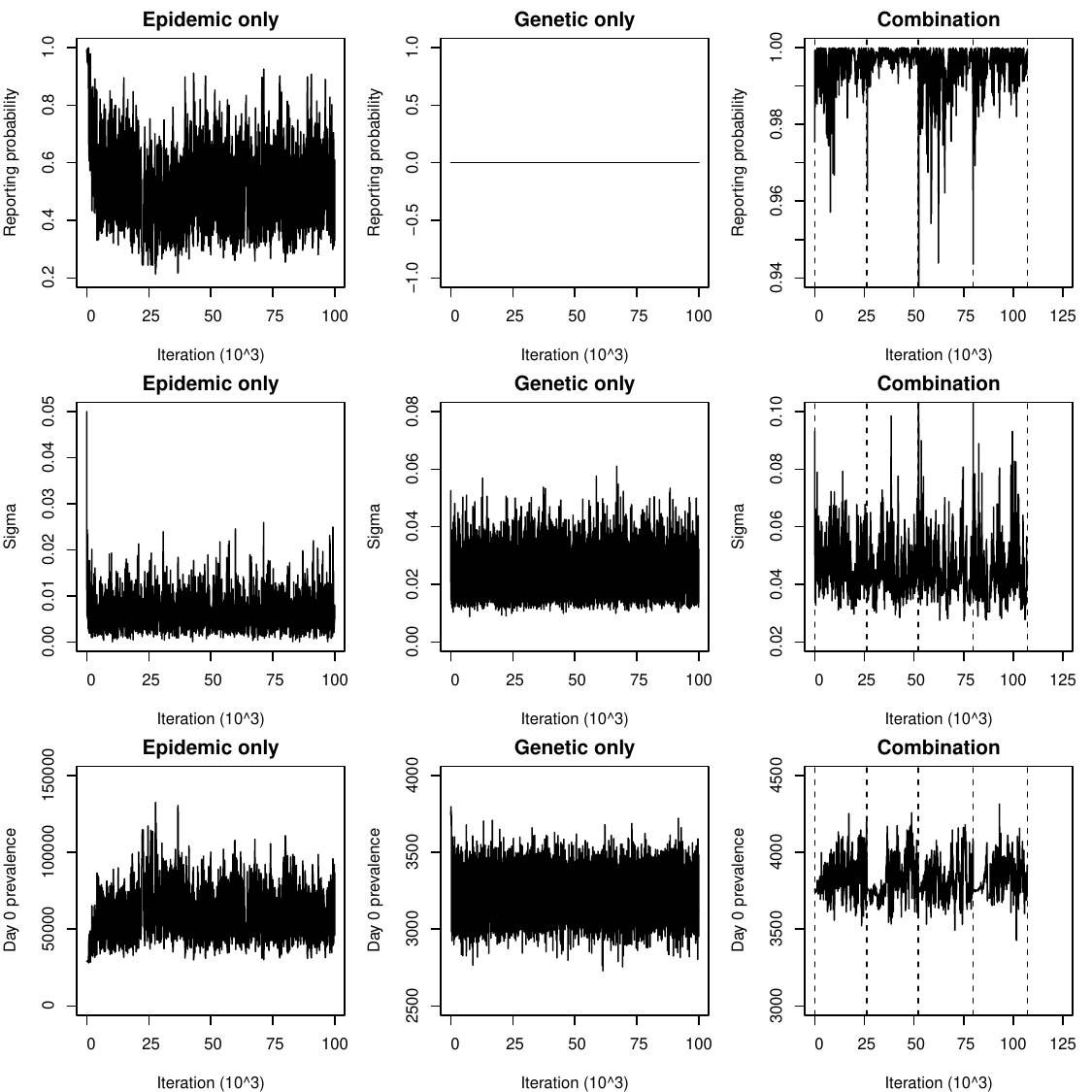}
\caption{Trace plots for $\rho$, $\sigma$ and $X_{0}$ for the PMMH algorithm run on the HIV-1 data set with epidemic data only, genetic data only and combination data.}
\label{fig:hiv_trace}
\end{figure}

\begin{figure}[ht]
\centering
\includegraphics[width=\textwidth]{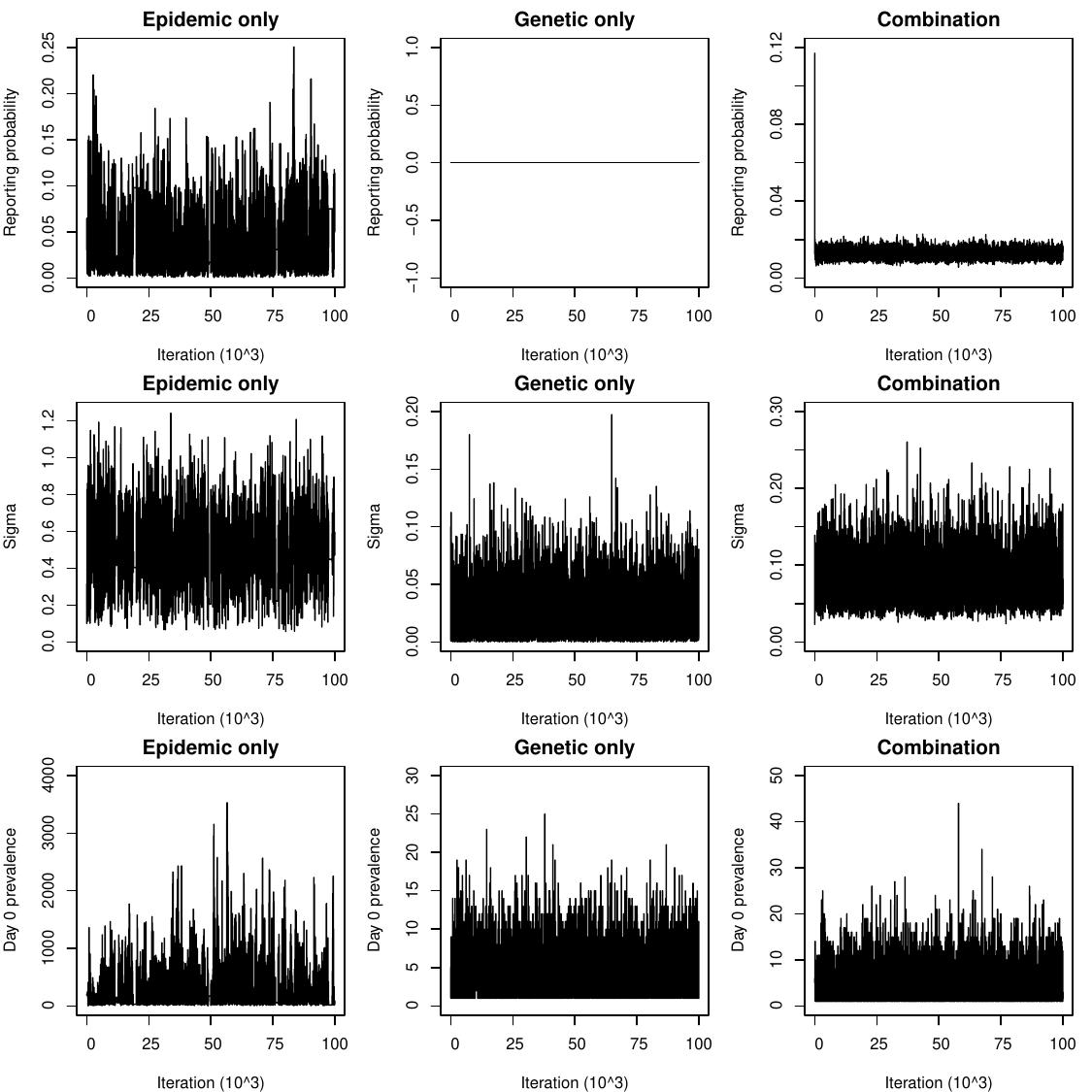}
\caption{Trace plots for $\rho$, $\sigma$ and $X_{0}$ for the PMMH algorithm run on the TB data set with epidemic data only, genetic data only and combination data.}
\label{fig:tb_trace}
\end{figure}

\end{document}